\documentclass[showpacs,amsmath,amssymb,aps,twocolumn,nofootinbib,notitlepage,rmp]{revtex4-1}

\usepackage{hyperref}

\usepackage{graphicx}
\usepackage{inputenc}
\usepackage[T1]{fontenc}

\usepackage{bbm}
\usepackage{lmodern,bm}

\usepackage[all,cmtip]{xy}

\newtheorem{theorem}{Theorem}

\newtheorem{proposition}{Proposition}
\newtheorem{example}{Example}
\newcommand{\h}[1]{\mathcal{#1}}
\newcommand{\R}{\mathbb{R}}  \let\Rl\R
  \let\Cx\C
\newcommand{\hi}{\mathcal{H}}

\newcommand{\lh}{\mathcal{L(H)}}
\newcommand{\tc}{\mathcal{T(H)}}
\newcommand{\hM}{{\h M}}
\def\instru{{\mathfrak I}}
\def\cbnorm#1{\Vert{}#1\Vert_{\rm cb}}
\def\id{{\rm id}}

\newcommand{\tr}[1]{\mathrm{tr}\left( {#1} \right)}
\newcommand{\ip}[2]{\left\langle\,#1\,|\,#2\,\right\rangle}
\newcommand{\kb}[2]{|#1\,\rangle\langle\,#2|}

\newcommand{\no}[1]{\left\|#1\right\|}
\def\idty{{\bf 1}}

\newcommand{\sfa}{{\sf A}}
\newcommand{\sfb}{{\sf B}}
\newcommand{\sfc}{{\sf C}}
\newcommand{\sfd}{{\sf D}}
\newcommand{\sfe}{{\sf E}}
\newcommand{\sff}{{\sf F}}
\newcommand{\sfg}{{\sf G}}
\newcommand{\sfh}{\mathsf{H}}
\newcommand{\sfm}{{\sf M}}
\newcommand{\sfq}{\mathsf{Q}}
\newcommand{\sfp}{\mathsf{P}}

\newcommand{\sfz}{\mathsf{Z}}

\newcommand{\bor}{\bm{r}}
\newcommand{\boa}{\bm{a}}
\newcommand{\bob}{\bm{b}}
\newcommand{\boc}{\bm{c}}

\newcommand{\bosig}{{\boldsymbol\sigma}}
\newcommand{\inpr}[2]{{#1}\cdot{#2}}

\newcommand{\stfm}{\mathfrak I}

\newcommand{\fii}{\varphi}
\newcommand{\veps}{\varepsilon}
\newcommand{\eps}{\epsilon}

\newcommand{\epno}{\eps_{\text{\sc no}}}
\newcommand{\etno}{\eta_{\text{\sc no}}}

 \def\swap{\mathbb F}

\def\im{\Im m}

\def\mitem[#1]{\par\noindent{\it#1}\/.\\ } 

\usepackage{color}

\begin{document}

\title{Colloquium: Quantum root-mean-square error and  measurement uncertainty relations}

\author{Paul Busch}
\email{paul.busch@york.ac.uk}
\affiliation{Department of Mathematics, University of York, York, YO10 5DD, United Kingdom}

\author{Pekka Lahti}
\email{pekka.lahti@utu.fi}
\affiliation{Turku Centre for Quantum Physics, Department of Physics and Astronomy, University of Turku, FI-20014 Turku, Finland}

\author{Reinhard F. Werner}
\email{reinhard.werner@itp.uni-hannover.de}
\affiliation{Institut f\"ur Theoretische Physik, Leibniz Universit\"at, D-30167 Hannover, Germany}

\date{October 9, 2014}

\begin{abstract}
Recent years have witnessed a controversy over Heisenberg's famous error-disturbance relation.  Here we resolve the conflict by way of an analysis of the possible conceptualizations of measurement error and disturbance in quantum mechanics. We discuss two approaches to adapting the classic notion of root-mean-square error to quantum measurements.  One is based on the concept of noise operator; its natural operational content is that of a mean deviation of the values of two observables measured jointly, and thus its applicability is limited to cases where such joint measurements are available. The second error measure quantifies the differences between two probability distributions obtained in separate runs of measurements and is  of unrestricted applicability. We show that there are no nontrivial unconditional joint-measurement bounds for {\em state-dependent errors}  in the conceptual framework discussed here, while Heisenberg-type measurement uncertainty relations for state-independent errors have been proven.

\end{abstract}


\maketitle
\tableofcontents

\section{Introduction}

In the past ten years, a  growing number of theoretical and experimental studies have claimed to challenge Heisenberg's uncertainty principle (e.g., \cite{Ozawa04a,Branciard2013} and \cite{Erh12,Roz12,Baek13,Ozawa-etal2014,Branciard2014}). Given the popular status of that fundamental principle, it is not surprising that these reports have created a considerable furore in popular science media and national newspapers across the world. 
While the challenge is ultimately unfounded (as will be shown here), it has helped to focus the attention of quantum physicists on a longstanding, important open problem:  to be sure, what is under debate is not the textbook version of Heisenberg's uncertainty relation that describes a trade-off between the standard deviations of the distributions of two observables in any given quantum state. Rather, the challenge is directed at another facet of Heisenberg's principle, the error-disturbance relation and, {\em a fortiori}, the joint measurement error relation. 

Perhaps surprisingly, in nearly ninety years of quantum mechanics, Heisenberg's celebrated ideas on quantum uncertainty have, to our knowledge, never been subjected to  direct experimental tests. This fact becomes less astonishing if one considers that neither Heisenberg nor, until rather recently, anyone else has laid the grounds to such experimental testing by providing precise formulations of error-disturbance relations and, more generally, joint measurement error relations. Ultimately, the reason for this omission lies in the fact that the conceptual tools for the description of quantum measurements had not been developed in sufficient generality until a few decades ago. Thus, for a long time research on the joint measurement problem was restricted to model investigations and case studies, and it was not until the late 1990s that the first general, model-independent formulations of measurement uncertainty relations were attempted.\footnote{For a review of this development we refer the interested reader  to \cite{BuHeLa07}.} Since then, in apparent contradiction to the alleged refutations of Heisenberg's principle, rigorous Heisenberg-type measurement uncertainty relations have in fact been deduced as consequences of quantum mechanics.  

The primary aim of this work is to explain the conceptual difficulties in defining appropriate quantifications of measurement error and disturbance needed for the formulation of such relations, and to describe how these difficulties have been overcome. As a byproduct we will see how the apparent conflict over Heisenberg's principle is resolved. It can be expected that this conceptual advance provides a firm basis for future investigations into harnessing quantum uncertainty for applications in quantum cryptography and quantum metrology.

The claim of a violation of Heisenberg's principle could only ever arise due to the informality of Heisenberg's own formulations. He gave only heuristic semi-classical derivations of his error-disturbance relation, which he expressed symbolically as
\begin{equation}\label{eqn:H-UR}
p_1\,q_1 \sim h.
\end{equation}
Here $q_1$ stands for the position inaccuracy and $p_1$ for the momentum disturbance, which Heisenberg identified with the spreads of the position and momentum distributions in the particle's (Gaussian) wave function after an approximate position measurement. 

Given the vagueness in Heisenberg's formulations of his uncertainty ideas, it is not clear what an appropriate rigorous formulation and generalization of Heisenberg's measurement uncertainty principle should look like. Rather than dwelling on historic speculations, we propose to take inspiration from Heisenberg's intuitive ideas and ask the question whether and to what extent quantum mechanics imposes limitations on the approximate joint measurability of two incompatible quantities. To give due credit to Heisenberg, we propose to call such limitations {\em Heisenberg-type measurement uncertainty (or error-disturbance) relations} if they amount to stipulating bounds on the accuracies (or disturbances) of simultaneously performed approximate measurements of two (or more) incompatible quantities, where the bound is given by a measure of the incompatibility.

Heisenberg's principle is paraphrased in, for example, \cite{Ozawa04a} or \cite{Erh12} as the statement that the measurement of one quantity, $A$, disturbs another quantity, $B$, not commuting with $A$ in such a way that certain so-called ``root-mean-square'' measures of error $\epno(A)$ and disturbance $\etno(B)$ (to be defined below) obey the trade-off inequality
\begin{equation}\label{eqn:false-ur}
\epno(A)\,\etno(B)\ge\tfrac 12\bigl|\langle\psi|[A,B]\psi\rangle\bigr|.
\end{equation}
It seems that the first reference to this inequality as  ``Heisenberg noise-disturbance uncertainty relation'' appears in \cite{Ozawa03b}. 
According to \cite{Erh12}, Heisenberg proved this inequality in his landmark paper of 1927 \cite{Heisenberg1927} on the uncertainty relation. Such a proof cannot be found in \cite{Heisenberg1927}, nor is there a formulation in this generality in any of Heisenberg's writings; finally, he did not use any explicit definition for measures of error and disturbance -- certainly not those of $\epno,\etno$. Hence there is no good reason to attribute the inequality \eqref{eqn:false-ur} to Heisenberg. It is therefore rather odd to base the claim of a refutation of Heisenberg's principle on a relation (inequality \eqref{eqn:false-ur}) that is actually incorrect {\em according to quantum mechanics itself} given the definitions of $\epno,\etno$ chosen by the authors of that claim.

Ozawa \cite{Ozawa04a}, Hall \cite{Hall04} and  Branciard \cite{Branciard2013} formulated inequalities (which are not entirely equivalent but of similar forms) that are  (mathematically sound) corrections of  \eqref{eqn:false-ur}. These inequalities, which all involve the quantities $\epno,\etno$  in addition to standard deviations, allow for the product $\epno(A)\etno(B)$ to be small and even zero without the commutator term on the right-hand side vanishing. A number of experiments have confirmed  the inequalities \cite{Erh12,Roz12,Weston-etal2013,Baek13,Ozawa-etal2014,Branciard2014}.

The definitions of the  quantities $\epno$ and $\etno$  in \eqref{eqn:false-ur} seem innocuous at first sight as they are based on the time-honored concept of the {\em noise operator}, which has a long history in the field of quantum optics, notably the quantum theory of linear amplifiers. 
Nevertheless, as we will show, $\epno$ and $\etno$ are problematic as  quantum generalizations of Gauss' root-mean-square deviations and hence their utility as estimates of error and disturbance is limited. 

In contrast, we will give here an extension of the concept of root-mean-square (rms) error that remains applicable without constraint in quantum mechanics. Our definition is based on the general representation of an observable as a positive operator valued measure, which is central to the modern quantum theory of measurement; as we will see, the observable-as-operator perspective underlying the noise operator approach has a rather more limited scope and can lead to conceptual problems if not applied judiciously.

Our measure of error 
obeys measurement uncertainty relations of the form
\begin{equation}\label{eqn:BLW-UR}
\Delta(Q)\,\Delta(P)\ge \frac\hbar2,
\end{equation}
which we have  proven in \cite{BLW2013c,BLW2013b} for canonically conjugate pairs of observables such as position and momentum. We emphasize that  $\Delta(A)$ is a state-independent measure of error and is not to be confused with the standard deviation  of an observable $A$ in a state $\rho$. We will also use the same concept for  qubit observables and review a form of {\em additive} trade-off relations for errors and for error and disturbance, with a non-trivial tight bound that is a measure of the incompatibility of the observables to be approximated; this new relation, presented in \cite{BLW2014}, can be tested in qubit experiments of the types reported in \cite{Erh12} and \cite{Roz12}. 

The paper is organised as follows. We begin with a brief discussion of the problem of conceptualizing measurement error and  disturbance in quantum mechanics (Section \ref{sec:conecpt-e-d}).
Here we draw attention to an important distinction between two perspectives on error and disturbance that relate to different physical purposes: on the one hand one may be interested in the interplay between the accuracy of a measurement performed on a particular state and the disturbance that this measurement imparts on the state; on the other hand there is a need to characterize the quality of a measuring device with figures of merit that apply to any input state. The work of Ozawa and Hall and of the experimental groups testing inequality \eqref{eqn:false-ur} and its generalizations is primarily concerned with the first type of task while our focus is mainly on the second. 

Another distinction to be addressed in Section \ref{sec:conecpt-e-d} concerns the purpose of error analysis: one may either be interested in the mean deviation of {\em values} or in a comparison of {\em distributions}.
The former kind of error measure is only applicable  in the restricted range of situations where quantum mechanics permits the joint measurability of the observables to be compared, whereas the latter is always applicable. The noise operator based measure is appropriately interpreted as a measure of the first type, and is therefore of limited use in quantum mechanics.

We then review the relevant elements of the language of quantum measurement theory (Section \ref{sec:language}). Next we recall the definitions of the noise operator based measures of error and disturbance (Sec.~\ref{sec:noise}) and present our alternative definitions based on a measure of distance between probability measures known as the Wasserstein 2-deviation (Sec.~\ref{sec:distance}). 
In Section \ref{sec:critique} we compare the quantities  $\epno,\etno$ with our distribution deviation measures, highlighting their respective merits and limitations. The inadequacy of the quantities $\epno,\etno$  as measures of error and disturbance for an {\em individual state} will be seen to be particularly striking in the qubit case. The analysis in this section will reveal in which circumstances and to what extent the quantities $\epno,\etno$ can be used as estimates of error and disturbance.

Finally  we review some formulations of the uncertainty principle that have been proven as rigorous consequences of quantum mechanics (Sec.~\ref{sec:theorems}). Among these are structural theorems describing measurement limitations and some forms of error-disturbance relations that can be considered to be in the spirit of Heisenberg's  ideas.

The paper concludes with a brief summary and survey of recent work on alternative formulations of measurement uncertainty relations inspired by the controversy over Heisenberg's principle (Sec.~\ref{sec:final}).

\section{The task of conceptualising error and disturbance}\label{sec:conecpt-e-d}

Here we consider how one should  define, say, the position error and momentum disturbance in measurement schemes such as, for instance, Heisenberg's microscope setup. The error $\Delta(A)$ of an approximate measurement of some observable $A$ clearly refers to the comparison of data obtained from two experiments, namely the given approximate measurement and an accurate reference measurement, so $\Delta(A)$ is a quantity comparing two measuring devices, assessing how much one fails to match the performance of the other.

A meaningful error analysis in an experiment requires that the proposed measure of error relates to the actual data obtained in the experiments to be compared; more explicitly, we hold that the following two requirements are necessary for any good error measure: 
\begin{itemize}
\item[(a)] an error measure is a quantification of the differences between the target observable and the approximator observable being measured; in particular it should correctly indicate cases where the target and approximating observables do agree, and where they do not;
\item[(b)] the error  can be estimated from the data obtained in the experiment at hand and an ideal reference measurement of the target observable. 
\end{itemize}

\subsection{Measurement error: comparing values or distributions?}

At this point it is necessary to reflect on the possibilities of implementing such an experimental error analysis. In classical physics it is common practice to test and calibrate the performance of a new measuring device by comparing its outputs  with those obtained in a highly accurate standard reference measurement. The mean error of the approximate measurement $C$ can then be defined as the root-mean-square (rms) deviation of its outcomes $c_k$ from the ``true value'', $a$, of the observable $A$ to be estimated, that is, symbolically, $\langle(c_k-a)^2\rangle^{1/2}$.

In quantum physics, it is only in the exceptional case of eigenstates that a quantity has a precise, definite value that could be revealed by an accurate measurement. If one does not want to restrict the assessment of the quality of a measurement as an approximation of a given observable to its eigenstates, one may consider calibrating the device by performing an accurate reference measurement jointly with the given measurement to be assessed. In this way one obtains value pairs, $(a_k,c_k)$, and as a substitute for the unknown or imprecise ``true'' value one can use the $A$ measurement values as reference for an error estimate, thus defining the {\em value comparison error} as the root-mean-square (rms) {\em value deviation}, $\langle(c_k-a_k)^2\rangle^{1/2}$.

However, the target observable $A$ and the observable $C$ measured to approximate it may not, in general, be compatible, so that a joint measurement will not be feasible. Therefore the value deviation concept is not universally applicable. Moreover, even in cases where $A$ and $C$ are compatible, the rms value deviation does not merely represent random noise and systematic errors inherent in the performance of the measuring device for $C$, but also encompasses preparation uncertainty of $A$ and correlations in the joint values of $A$ and $C$.

In order to find a uinversally applicable measure of error for quantum measurements, one must therefore look for an alternative approach. Since the signature of an observable is the totality of its statistics for all states, a viable method that offers itself is to apply the reference measurement and the approximate measurement to {\em different} ensembles of objects in the same state;  one can then compare the two measurement outcome {\em distributions}. This method may be referred to as {\em distribution error} estimation.

We will see that the definition of error  used by Ozawa and collaborators are appropriately understood as formal extensions of the value comparison error concept; they must therefore be expected to be of limited use. Examples given below will demonstrate that where they fail to meet  requirement (b), they also become unreliable and so fall short of (a) as well. Our alternative error measure is an instance of the distribution error method.

For the {\em disturbance} $\Delta(B)$ of an observable  $B$ in a measurement of $A$ (such as the disturbance of the momentum in a microscope observation) we face the same issues. One has to allow for the possibility that the momenta before and after the measurement interaction do not necessarily commute, so the difference cannot be determined  by comparing individual values to be obtained in joint measurements. In contrast, it is always possible to compare the {\em distribution} of the measured momenta after the position measurement with the {\em distribution} of an  accurate momentum measurement performed directly on the same input state.

\begin{figure}
\begin{center}
\includegraphics[width=8cm]{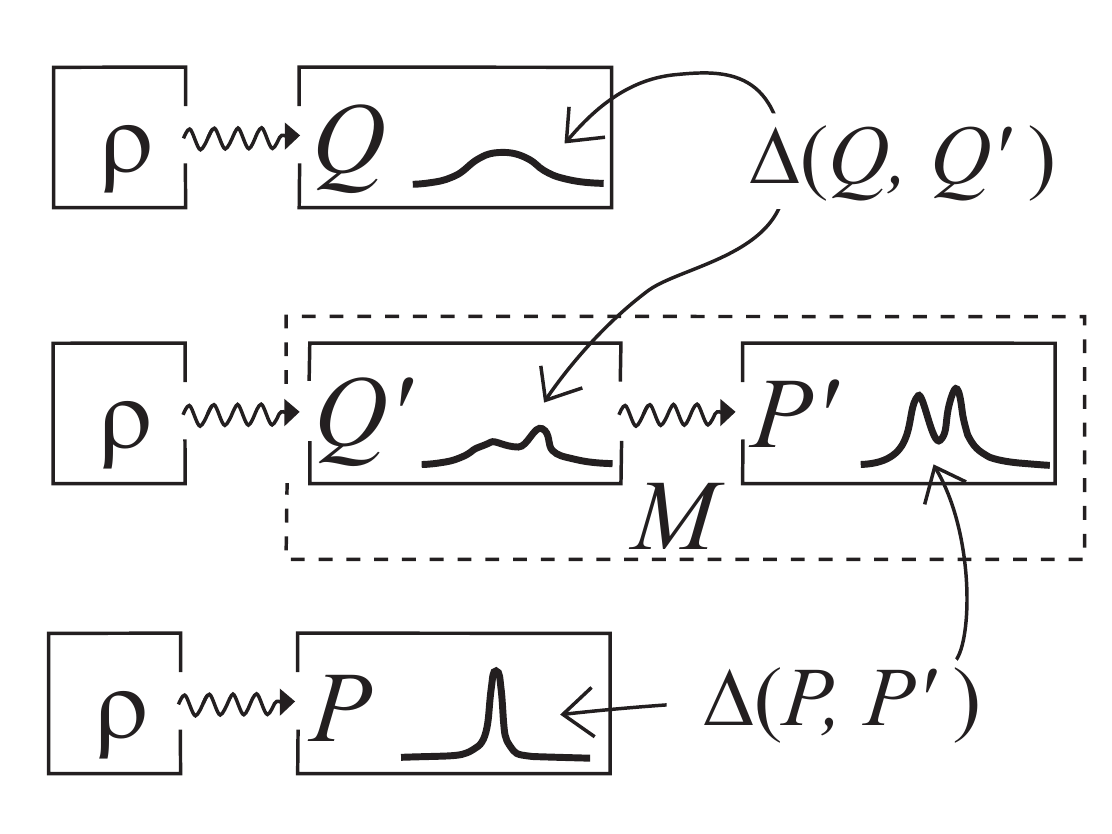}
\end{center}
\caption{Comparison of experiments 
involved in an error-disturbance relation. The dotted box indicates that the sequential measurement consisting of first performing an approximate position and then an ideal momentum measurement can just be considered as a single approximate joint measurement. The joint measurement view thus restores the symmetry between position and momentum in uncertainty relations.}
\label{fig:boxes}
\end{figure}

This is precisely how we detect disturbance in other typical quantum settings. Consider, for example, the double slit experiment. Illuminating the slits enough to detect the passage of a particle through one or the other hole makes the interference fringes disappear.  Clearly the light used for the observation disturbs the particles, and the evidence for this is once again the change of the distribution on the screen. This is illustrated schematically in Fig.~\ref{fig:boxes}. 

\subsection{State-specific error vs. device figure of merit}

The problem of quantifying measurement error and disturbance may be approached in two distinct ways. First, one may be interested in the question of how close a given measurement device comes to realizing a good approximate measurement of some observable in a particular fixed state of the system. This question can be approached by defining {\em state-specific} error and disturbance measures. Such state-dependent measures would allow one to determine the imprecision that one has to accept in the measurement of some observable if it is required that the disturbance imparted on some other observable should be limited to a specified amount.

We have already seen that the notion of value-comparison error does not lend itself to being widely applicable to quantum measurements; thus it appears that one must take resort to using distribution comparison errors. However, state-dependent distribution comparison measures do not yield nontrivial joint measurement error bounds or error-disturbance trade-off relations, as shown in the following example.

Consider a perfectly accurate position measurement where the state change is given as a constant channel. For any given state $\rho$, one can choose the measurement such that the constant channel output state is identical to $\rho$;  then no disturbance of the state occurs, and any  error and disturbance measures that just compare distributions will have value zero.

For some time the only state-dependent error approach to formulating measurement uncertainty relations has been that of Ozawa \cite{Ozawa04a} and Hall \cite{Hall04}, which is based on the noise operator based quantities $\epno,\etno$. We will provide evidence showing that these quantities are only useful as error and disturbance measures for a limited class of measurements.  It follows that Ozawa's and other inequalities based on $\epno$ and $\etno$ cannot claim to be universally valid uncertainty relations -- these inequalities do admit an interpretation as error/disturbance trade-off relations for a limited class of approximate joint measurements only.

The second approach to quantifying measurement errors is one of interest to a device manufacturer, who would wish to specify a worst-case limit on the error and disturbance of a device; this would allow the customers to be assured of (say) an overall error bound that applies to {\em all} states they wish to measure. Such {\em device figures of merit} will thus be {\em state-independent} measures of error and disturbance.

There are (at least) two ways of obtaining state-independent error measures. The first is to define a state-dependent measure for all states and define {\em the worst case error} as the least upper bound of these numbers.
Alternatively, one can focus on a representative subset of states, namely, the (near-)eigenstates, and define  the mean or the worst-case error across these. Error measures obtained by the latter method will be called {\em calibration errors}.

Realistic measuring devices will not normally work on all input states; they have a {\em finite operating range}. For the purposes of the present paper we will mainly maintain the idealization of allowing arbitrary input states;
this is in line with the common idealized representation of observables like position and momentum as unbounded operators with an infinite range of possible values.
As mentioned, one way of taking into account the finite operating range is to consider calibration error measures. 

Measurement uncertainty relations for such overall errors and calibration errors were proven in \cite{Appleby1998a,Appleby1998b,Werner04,BuPe07} for various state-independent measures, and more recently in \cite{BLW2013b,BLW2013c,BLW2014} for a general family of error measures.  Some of these results will be reviewed in Section \ref{QPUR}.

\section{Operational Language of Quantum Mechanics}\label{sec:language}

We  review briefly the key tools of {\em operational quantum mechanics} (e.g., \cite{Davies,Holevo,Ludwig1983,QTM}) required for our analysis; these are: {\em observables} as positive operator valued measures; the description of state changes through measurements in terms of the notion of {\em instrument}; and the general concept of {\em measurement scheme}. We will also comment on the restrictive {\em observable-as-operator} point of view that is still predominant in the textbook literature but becomes problematic when adhered to in the modeling of approximate measurements and the search for measures of approximation errors.

\subsection{Observables}\label{sec:POVM}

In quantum mechanics, the states of a physical  system are generally represented as the positive trace-one operators, also called density operators, acting on the Hilbert space $\hi$ associated with the system. Any observable of the system is uniquely determined through the  distributions of measurement outcomes associated with the states $\rho$; thus an observable $\sff$ can be described as a map that associates a probability measure $\sff_\rho$ with every state, $\rho\mapsto \sff_\rho$, where $\sff_\rho$ is defined on the set $\Omega$ of outcomes, equipped with a $\sigma$-algebra of subsets $\Sigma$. The form of the distributions is automatically in accordance with the Born rule:  $\sff_\rho(X)=\tr{\rho \sff(X)}$. 
Here $\sff(X)$ is a positive operator for each $X\in\Sigma$ with $\sff(X)\leq\idty$ (such operators are called {\em effects}), and $X\mapsto \sff(X)$ the {\em normalized positive operator (valued) measure} (occasionally abbreviated POVM or POM) representing the observable $\sff$. 
The standard, {\em sharp} observable, given by a spectral measure, is included as a special case.

For any (measurable) scalar function $f$, one can define a unique linear operator $\sff[f]$ such that $\ip{\psi}{\sff[f]\psi}  =\int  f(x)\, \sff_\rho(dx)$ for all  $\rho=\kb\psi\psi$ with $\int |f|^2\,\sff_\rho(dx)<\infty$. In the case of measurements with real values ($\Omega=\Rl$) we follow a widespread abuse of notation by denoting functions $x\mapsto x^n$  by their values. 
Thus we can define the {\em moment operators} $\sff[x^n]$ of $\sff$ through the moments $\sff_\rho[x^n]=\int x^n\,\,\sff_\rho(dx)$  of the distribution $\sff_\rho$; with a slight abuse of notation we also
write $\langle\sff[x^n]\rangle_\rho\equiv\tr{\rho \sff[x^n]}$ for $\int x^n\,\,\sff_\rho(dx)$ whenever $\int x^{2n}\,\,\sff_\rho(dx)<\infty$.

If $\sff$ is a projection valued measure, then $\sff[x]$ alone determines this measure $\sff$ uniquely, and the domain of $\sff[x]$ consists of the vectors $\psi$ for which the square integrability condition $\int x^2 \,\ip{\psi}{\sff(dx)\psi}<\infty$ holds. $\sff$ is then the {\em spectral measure} of the selfadjoint operator $\sff[x]$.

If $A$ is a selfadjoint operator, we let $\sfa$ (or also $\sfe^A$) denote the unique spectral measure associated with $A$, so that $A=\sfa[x]=\sfe^A[x]$. Since the distinction between operator measures and operators is so crucial for the topic in question, we always use Sans Serif type letters like  $\sfa$ for observables (as measures) and Roman type letters for operators like $A$, even for sharp observables where $\sfa$ and $A$ are in one-to-one  correspondence with each other.

For a general POVM  $\sff$ the operator $\sff[x]$
does {\it not} determine the full probability distributions;  many different POVMs may have the same first moment operator, so it makes no sense to call this operator ``the observable''. 
Von Neumann's terminology (in which operators and observables are the same thing) is so deeply rooted 
in physics education, that it seems appropriate to elaborate once more on the difference between observables 
and their first moment operators, especially since the conflation  directly enters the definition of the quantities 
$\epno,\etno$. 

Even in the context of  projection valued observables alone, there is good reason to distinguish conceptually between the operator and its spectral measure. Indeed, there are situations where for two noncommuting observables $\sff$ and $\sfg$  the sum operator $H=\sff[x]+\sfg[x]$ is selfadjoint (or has a selfadjoint extension). It is then clear how to set up an experiment to determine the expectation $\tr{\rho H}$, namely by measuring $\sff$  on a part of the sample and $\sfg$ on the rest, and adding the expectation values. However, there are no ``outcomes'' $h\in\Rl$, which appear in this combined experiment, and no probability distribution associated with that operator. One has simply performed two incompatible measurements on different parts of a sample of equally prepared systems. In particular, there is no way to directly determine $\tr{\rho H^2}$ from the two measurements. 

If we follow the Rules of the Book, this is how we should do it: Compute the spectral measure $\sfh$ so that $H=\sfh[x]$. Then {\it invent a new experiment} in which this observable is measured. Next, measure this new observable on $\rho$, and compute the second moment of the statistics thus obtained. The problem is that we have no handle on how to design a measurement of the observable $\sfh$. The connection between $\sff$, $\sfg$, and $\sfh$ is, in fact,  so indirect, that a good part of most quantum mechanics textbooks is devoted to the simplest instances of  this task: Diagonalizing the sum of two non-commuting operators (namely kinetic and potential energy if $H$ is the Hamiltonian), each of which has a simple, explicitly known diagonalization. This problem is further underlined by a subtlety for unbounded operators: Even if the summands are both essentially selfadjoint on a common domain, their sum may fail to be so as well, so that the expectation of $H$ is well-defined but not the spectral resolution.

Since for a general (real) observable $\sff$ the second moment cannot be computed from the first, it is sometimes helpful to quantify the difference. We have $\sff[x^2]\geq \sff[x]^2$ in the sense that the {\it variance form} 
\begin{equation}\label{varform}
  V_\sff(\phi,\psi)=
  \int x^2\ip\phi{\sff(dx)\psi}-\ip{\sff[x]\phi}{\sff[x]\psi},
\end{equation}
defined for $\phi,\psi$ in the domain of $\sff[x]$, is non-negative for $\phi=\psi$ (see \cite{Werner1986,KLY2006}). Sometimes this extends to a bounded operator which we denote by $V(\sff)$, so $\ip\phi{V(\sff)\psi}= V_\sff(\phi,\psi)$. In particular, if $\sff[x]$ is selfadjoint, then $\sff[x^2]\geq \sff[x]^2$ on the domain of $\sff[x]$ and the difference operator $V(\sff)=\sff[x^2]-\sff[x]^2$, occasionally called the {\em intrinsic noise} operator, allows one to express the variance $\Delta(\sff_\rho)^2=\int(x-\int x\,\sff_\rho(dx))^2\,\sff_\rho(dx)$ of an observed probability distribution $\sff_\rho$  as a sum of two non-negative terms:
\begin{equation}\label{varsum}
  \Delta(\sff_\rho)^2
     =\tr{\rho V(\sff)}+ \Delta(\sfe^{\sff[x]}_\rho)^2.
\end{equation}
This shows that the distribution of the observable $\sff$ is always broader than the distribution of the sharp observable represented by $\sff[x]$ (assuming the latter is a selfadjoint operator), and the added noise is due to the intrinsic unsharpness of $\sff$ as measured by $V(\sff)$. It is worth noting that this equation presents a splitting of the variance of the probability distribution $\sff_\rho$ into two terms that are not accessible through the measurement of $\sff$: the term $\tr{\rho \sff[x]^2}$ cannot be determined from the statistics of $\sff$ in the state $\rho$ -- unless $\sff$ is projection valued, which is equivalent to $\sff[x]$ being selfadjoint and $\sff[x]^2=\sff[x^2]$, that is, $V(\sff)=0$. 

\begin{example}\label{convolution}\rm
Consider an observable on $\R$ of the convolution form $\mu*\sff$, with a fixed (real) probability measure $\mu$. Thus, $\mu*\sff$ is the unique observable defined by the map $\rho\mapsto\mu*\sff_\rho$, where the convolution $\mu*\nu$ of two (real) probability measures $\mu,\nu$ is the unique probability measure  defined via the product measure $\mu\times\nu$, 
\[
(\mu*\nu)(X)=(\mu\times\nu)(\{(x,y)\in\Rl^2\,|\, x+y\in X\}).
\]
For later use we note that $\Delta(\mu*\sff_\rho)^2=\Delta(\mu)^2+\Delta(\sff_\rho)^2$ and the intrinsic noise operator is the constant operator $V(\mu*\sff)=\Delta(\mu)^2\,\idty$ (with the obvious restrictions on the domains and assuming that $\Delta(\mu)<\infty$). ---
\end{example}

\subsection{Measurements}\label{sec:instruments}
There are two equivalent ways to model measurement-induced state changes. One can either use an ``axiomatic'' description starting from a set of minimal requirements imposed by the statistical interpretation of the theory. This leads to the definition of an {\em instrument}.\footnote{The concept of an instrument as an operation-valued measure was introduced by Davies and Lewis in the late 1960s \cite{Davies}. These authors did not explicitly stipulate the complete positivity of operations as part of the definition, a property that was already known to be a crucial feature required from the perspective of measurement theory (e.g., \cite{Kraus74, Davies,Kraus83}). Here we follow the practice introduced in \cite{Ozawa1984} of including complete positivity in the definition of an instrument.} Alternatively, one can work ``constructively'' and describe a {\em measurement scheme}  involving a unitary coupling  between the object and a measurement device and subsequent measurement of a pointer observable on the measuring device.\footnote{A modern presentation of this latter approach, which goes back to von Neumann \cite{vN32}, can be found, for instance, in \cite{QTM}.} That these approaches agree -- a consequence of the Stinespring dilation theorem -- makes the definition of the class of measurements very canonical.

Given a physical system with Hilbert space $\hi$, an {\em instrument} $\instru$ describes all the possible output states of a measurement  conditional on the values from an outcome space $\Omega$; it is thus a collection of completely positive maps on the trace class, $\instru(X):\tc\to\tc$, labeled by  the (measurable) sets $X\subseteq \Omega$ of outcomes, such that for each input state $\rho$ the map $X\mapsto \tr{\instru(X)(\rho)}$ is a probability measure. The interpretation is that $\tr{\instru(X)(\rho)B}$ is the probability for a measurement result $x\in X$ in conjunction with the `yes' response of some effect $B\in\lh$ ($0\leq B\leq\idty$) after the measurement. When we ignore the outcomes there is still a disturbance of the input state $\rho$, represented by the {\em channel} $\rho\mapsto\instru(\Omega)(\rho)$. Alternatively, we may choose to ignore the system after the measurement, setting $B=\idty$ in the probability expression, and obtain an observable $\sff$ on $\Omega$ via\footnote{Here we are using the notation $\instru^*$ for the dual instrument to $\instru$, defined via the relation $\tr{\instru(X)(\rho) B}=\tr{\rho \instru(X)^*(B)}$, required to hold for all $\rho,X,B$.} 
\begin{equation}
\tr{\rho \sff(X)}=\tr{\instru(X)(\rho)}=\tr{\rho\,\instru(X)^*(\idty)}.
\end{equation}
It is a simple observation that for any observable $\sff$ there is an instrument $\instru$ such that $\tr{\rho \sff(X)}=\tr{\instru(X)(\rho)}$ and that the association $\instru\mapsto \sff$ is many-to-one. For later reference we note the class of instruments with constant channel associated with an observable $\sff$ and a fixed state $\rho_0$, where
\begin{equation}\label{eqn:const-instru}
\instru^\sff_{\rho_0}(X)(\rho)=\tr{\rho \sff(X)}\rho_0.
\end{equation}
The disturbance exerted by this type of instrument on any observable $\sfb$ has the effect of turning $\sfb$ into a trivial observable $\sfb'$:
\begin{equation}\label{eqn:const-disturb}
\tr{\rho \sfb'(Y)}=\tr{\rho\instru^\sff_{\rho_0}(\Omega)^*\bigl(\sfb(Y)\bigr)}=\tr{\rho_0\sfb(Y)}
\end{equation}
for all $Y$, so that $\sfb'(Y)=\sfb_{\rho_0}(Y)\idty$.

A {\em measurement scheme} $\hM$ comprises a probe system in a fixed initial state $\sigma$ from its Hilbert space $\mathcal{K}$, a unitary map $U$ representing the coupling of object and probe that enables the information transfer, and a probe observable $\sfz$ representing the pointer reading.\footnote{The probe observable can always be assumed to be a sharp observable so that we may also refer to $Z=\sfz[x]$ as the probe observable.} This is connected with the notion of instrument and the observable $\sff$ by 
\begin{eqnarray}\label{instruScheme}
  \tr{\instru(X)(\rho)B}&=&\tr{(\rho\otimes\sigma)U^*(B\otimes \sfz(X))U} \\
  \tr{\rho\, \sff(X)}&=&\tr{(\rho\otimes\sigma)U^*(\idty\otimes \sfz(X))U}. \label{obsScheme}
\end{eqnarray}
In the first case, these formulas show that each measurement scheme $\h M$ defines an instrument $\instru$ and the accompanying observable $\sff$. The converse result is obtained from the Stinespring dilation theorem for completely positive instruments. We summarize this fundamental connection in a theorem. (To the best of our knowledge, the first explicit proofs of these results in this generality is given in \cite{Ozawa1984}.)

\begin{theorem}\label{QTM}
Every measurement scheme $\h M$ determines an instrument $\instru$ and an observable $\sff$ through (\ref{instruScheme}) and (\ref{obsScheme}). Conversely, for each instrument $\instru$ and thus observable $\sff$, there exist measurement schemes $\h M$ implementing them, in the sense that (\ref{instruScheme}) and (\ref{obsScheme}) hold.
\end{theorem}

\subsection{Sequential and joint measurements}

A {\em sequential measurement scheme} for two observables $\sff,\sfg$ with respective value spaces $\Omega_1,\Omega_2$ is defined via formula (\ref{instruScheme}) when the effects $B$ are chosen to be those of an  observable $Y\mapsto \sfg(Y)$; then for any $X\subset \Omega_1,Y\subset \Omega_2$,
\begin{equation}\label{sinstruScheme}
  \tr{\instru(X)(\rho)\sfg(Y)}
  =\tr{(\rho\otimes\sigma)U^*(\sfg(Y)\otimes \sfz(X))U},
\end{equation}
defines a sequential {\em biobservable}  $(X,Y)\mapsto \sfe(X,Y) = \instru(X)^*(\sfg(Y))$, with the probabilities of pair events ({\em biprobabilities}) given as
\begin{equation}\label{seqbiobservable}
\tr{\rho \sfe(X,Y)}= \tr{\rho\, \instru(X)^*(\sfg(Y))}.
\end{equation}
The two marginal observables $\sfe_1,\sfe_2$ are 
\begin{eqnarray}\label{A-marginal}
\sfe_1(X)=\sfe(X,\Omega_2)&=& \instru(X)^*(\idty)=\sff(X),\\
\sfe_2(Y)=\sfe(\Omega_1,Y) &=&\instru(\Omega_1)^*(\sfg(Y)) =:\sfg'(Y).\label{B-marginal}
\end{eqnarray}
This shows that the first marginal observable is the observable $\sff$   measured first by $\h M$, whereas the second marginal observable $\sfg'$ is a {\em distorted} version of the second measured observable $\sfg$, the distortion being a result of the influence of $\h M$.

There is an important special case.
\begin{proposition}\label{prop:productform}
If one of the marginal observables of a sequential biobservable $\sfe$ is projection valued, then 
\begin{equation}\label{productform}
\sfe(X,Y)=\sfe_1(X)\sfe_2(Y)=\sfe_2(Y)\sfe_1(X)
\end{equation}
for all $X,Y$.
\end{proposition}
For a  proof of this presumably well-known result we quote \cite[Theorem 1.3.1, p.91]{Ludwig1983} together with \cite[Lemma 1]{KLS2009}.

We say that two observables $\sff$ and $\sfg$ (with value sets $\Omega_1$ and $\Omega_2$) are {\em jointly measurable} if there is a measurement procedure that reproduces the statistics of both in every state; that is, there exist a measurement scheme $\h M$ and (measurable) pointer functions $f$ and $g$ such that
\begin{eqnarray}\label{jointE}
 \tr{\rho\, \sff(X)}&=&\tr{(\rho\otimes\sigma)U^*(\idty\otimes \sfz(f^{-1}(X)))U},\\
 \tr{\rho\, \sfg(Y)}&=&\tr{(\rho\otimes\sigma)U^*(\idty\otimes \sfz(g^{-1}(Y)))U}.\label{jointG} 
\end{eqnarray}
If $\sfm$ is the observable defined by $\h M$ through (\ref{obsScheme}), then $\sff(X)=\sfm(f^{-1}(X))$ and $\sfg(Y)=\sfm(g^{-1}(Y))$, that is, $\sff$ and $\sfg$ are functions of $\sfm$. 
An alternative definition of joint measurability requires the existence of a {\em joint observable} for $\sff$ and $\sfg$, that is, an observable $\sfe$ defined on the ($\sigma$-algebra of subsets of $\Omega_1\times\Omega_2$ generated by the) product sets $X\times Y$ such that $\sff$ and $\sfg$ are its marginal observables
\begin{equation}
\sff(X)=\sfe_1(X) \quad{\rm and}\quad 
\sfg(Y)=\sfe_2(Y).
\end{equation}
These two notions of joint measurability are known to be equivalent. If $\sff$ and $\sfg$ have a joint observable $\sfe$, they are also jointly measurable. The converse result, that  the biobservable $(X,Y)\mapsto \sfm(f^{-1}(X)\cap g^{-1}(Y))$ extends to a  (unique) joint observable of its marginal observables holds, in particular,  in the case of observables on $\R$. This is a consequence of a more general statement proven e.g. in \cite[Theorem 1.10, p. 24]{BCR1984}. Hence, for any two observables on $\R$ the following three conditions are equivalent: they have a biobservable; they have a joint observable; they are functions of a third observable.

\section{Noise-operator based error}\label{sec:noise}

We now review the definition of the noise-based quantities $\epno,\etno$ and associated uncertainty relations. 

\subsection{Definitions}

Consider a measurement scheme   $\hM=(\mathcal K,\sigma,Z,U)$ as an approximate measurement  of a sharp observable $A=\sfa[x]$. We will denote by $\sfc$ be the observable determined by $\hM$.  Instead of seeking a measure that quantifies the difference between  the distributions $\sfc_\rho$ and $\sfa_\rho$, the noise-operator approach defines the error in approximating  $\sfa$ with $\h M$ in a state $\rho$ via
\begin{equation*}
  \epno(A,\hM,\rho)^2=\tr{(\rho\otimes\sigma)\bigl(U^*(\idty\otimes Z)U - A\otimes\idty\bigr)^2}.
\end{equation*}
This expression is usually justified with an appeal to classical analogy (e.g., \cite{Ozawa-etal2014}), where it would represent the root-mean-square deviation between the values of two simultaneously measured random variables.

The state change caused by $\hM$ is described by the associated instrument via the  channel $\rho\mapsto\stfm(\R)(\rho)$;
this entails that the initial distribution $\sfb_\rho$ of any other sharp observable $\sfb$ is changed to $\sfb_{\stfm(\R)(\rho)}\equiv \sfb'_\rho$. Again, instead of comparing the distributions $\sfb_\rho$ and $\sfb'_\rho$, the noise-operator approach takes the disturbance caused by $\hM$ on $\sfb$ in a state $\rho$ to be quantified by
\begin{equation*}
  \etno(B,\hM,\rho)^2=\tr{(\rho\otimes\sigma)\bigl(U^*(B\otimes \idty)U - B\otimes\idty\bigr)^2},
\end{equation*}
where $B$ is the unique selfadjoint operator defining $\sfb$.

\subsection{Historic comments}

With the notation $\epno$, $\etno$ we indicate the underlying {\em observable-as-operator} point of view. These quantities are defined via expectations of the square of an operator that is the difference of an input and output operator. We will refer to $\epno,\etno$ as {\em NO-error} and  {\em NO-disturbance}, since they are modeled after the concept of  {\em noise operator} in quantum optics,
which was formalized by Haus and Mullen in 1962 \cite{HausMullen1962} as the difference of the operators representing the signal and output  (for some useful reviews, see \cite{Haus2004,YaHaus1986,Clerk2010}).

The use of the noise operator in the modeling of quantum measurement error can be traced to the seminal work of Arthurs and Kelly \cite{AK65}, which was elaborated further by Arthurs and Goodman \cite{AG88}. The quantity $\epno$ appears there as an auxiliary entity in the derivation of generalized preparation uncertainty relations for the output distributions in a simultaneous measurement of conjugate quantities that reflect the presence of the inevitable fundamental measurement noise. It is of interest to note that in these works, no independent operational meaning is expressly assigned to $\epno$, and the inequality 
\begin{equation}\label{eqn:AK-unbiased}
\epno(A,\hM,\rho)\,\epno(B,\hM,\rho)\ \ge\ \tfrac 12\bigl|\tr{\rho [A,B]}\bigr|
\end{equation}
for a joint approximate measurement of two observables $A,B$
is deduced under the assumption of unbiased approximations. Somewhat later, rigorous proofs  of this inequality for unbiased measurements were given by Ishikawa \cite{Ishikawa1991} and Ozawa \cite{Ozawa1991}.

The approach of Arthurs and Kelly was taken up  by Appleby \cite{Appleby1998a},  who used it to  formulate various kinds of joint measurement error and disturbance relations.  He clearly recognized  that inequalities  of the form \eqref{eqn:false-ur}, \eqref{eqn:AK-unbiased}  are bound to fail for state-dependent measures; accordingly he proceeded to deduce state-independent measurement uncertainty relations for generic joint measurements of position and momentum \cite{Appleby1998b}, using the suprema of $\epno,\etno$ over all states. He also generalized these relations to approximate measurements with finite operating range (see Subsection. \ref{QPUR}).

\subsection{Ozawa's inequality and generalizations}

For the  numbers $\epno,\etno$ Ozawa derives the inequality
\begin{equation}\label{eqn:Ozawa-inequality}
\begin{split}
& \epno(A,\h M,\rho)\,\etno(B,\h M,\rho)
\,+\,\epno(A,\h M,\rho)\Delta(\sfb_\rho)\\ 
&\quad + \Delta(\sfa_\rho)\etno(B,\h M,\rho)
\, \geq\ \tfrac 12\bigl|\tr{\rho [A,B]}\bigr|\,,
\end{split}
\end{equation}
which is proposed as   a universally valid error-disturbance relation. There is a  corresponding joint measurement error relation where $\hM$ is an approximate joint measurement of $A$ and $B$; this is obtained by substituting $\epno(B,\h M,\rho)$ for $\etno(B,\h M,\rho)$.

Ozawa's inequality has recently been strengthened by Branciard \cite{Branciard2013} for the case of pure states 
$\rho=\kb\fii\fii$
(here we are using the simplified notation $\epno(A,\h M,\rho)\equiv{\epno}(A)$, etc.):
\begin{align}
&{\epno}(A)^2\Delta(\sfb_\rho)^2+\epno(B)^2\Delta(\sfa_\rho)^2\nonumber\\
&+2\sqrt{\Delta(\sfa_\rho)^2\Delta(\sfb_\rho)^2-\tfrac 14|\langle[A,B]\rangle_\fii|^2}\,\epno(A)\epno(B)\nonumber\\
&\qquad\qquad\qquad\qquad\qquad\qquad\quad\ge \tfrac 14|\langle[A,B]\rangle_\fii|^2.\label{eqn: Branciard}
\end{align}
This inequality is in fact tight: for any $A,B,\rho=\kb\fii\fii$, there are measurements $\hM$ for which equality is achieved. 

As noted earlier,  variations of Ozawa's inequality based on the quantity $\epno$ have been proposed, notably in \cite{Hall04} and  \cite{Weston-etal2013}. Branciard \cite{branciard2013deriving} has shown that these three types of (inequivalent) inequalities can be obtained as special cases of his own.

\section{Distribution errors}\label{sec:distance}

\subsection{Distance between distributions}
As noted earlier, quantum measurement errors cannot in general be determined as value deviations by performing the approximate measurement jointly with an accurate control measurement {\em on the same system}. But they can be estimated as distribution deviation measures, namely, by comparing the actual statistics with those of an {\em independent} (and ideally accurate) reference measurement of the target observable on a separate ensemble of systems prepared in the same state. When the state is fixed, the comparison thus amounts to an evaluation of the difference between two probability distributions. Therefore, the key to a definition of the quality of a measurement, as compared to an ideal one, lies in finding a {\em measure of distance} between two probability measures. 

For a general outcome space $\Omega$ there are many ways of doing this, just as there are many ways of defining a metric on $\Omega$. For uncertainty relations, however, we want, for instance, the distance between position measurements to be in physical length units. This is a requirement of scale invariance, and also fixes the metric on $\Omega$ to be the standard Euclidean distance. A similar consideration is encountered in the definition of the ``spread'' of a probability distribution, as needed in the preparation uncertainty relation. The conventional root-mean-square deviation clearly has the right units, but so does a whole class of the so-called power-$\alpha$ means. Instead of developing the general theory (cf.~\cite{BLW2013b}) we consider here only  the case of $\alpha=2$ and $\Omega=\Rl$, equipped with the Euclidean distance $D(x,y)=|x-y|$.

Identifying a fixed point $y\in\Rl$ with the point measure $\delta_y$ concentrated at $y$,  the root-mean-square deviation
\begin{equation}\label{Dpoint}
\Delta(\mu,\delta_y)=\left(\int |x-y|^2\,\mu(dx)\right)^{\frac 12}
\end{equation}
is a measure for the deviation of a probability measure $\mu$
 from the point measure $\delta_y$. In particular, $\Delta(\delta_x,\delta_y)=D(x,y)$, which further emphasizes the intimate connection of the deviation with the underlying metric structure of $\Omega$, here  $\Rl$.  The standard  deviation is then
\begin{equation}\label{Delalpha}
 \Delta(\mu) =\inf\{ \Delta(\mu,\delta_y)\,|\, y\in\Rl\},
\end{equation}
with the minimum obtained for $y=\mu[x]$ (if finite).

The  deviation (\ref{Dpoint}) can readily be extended to any pair of probability measures $\mu,\nu$ using their {\em couplings}, that is, probability measures $\gamma$ on $\Rl\times\Rl$ having $\mu,\nu$ as the (Cartesian) marginals. Given a coupling  $\gamma$ between $\mu$ and $\nu$ one may  define 
\begin{equation}\label{Dcoupling}
 \Delta^\gamma(\mu,\nu)=\left(\int |x-y|^2\,\gamma(dx,dy)\right)^{\frac 12},
\end{equation}
as a deviation of $\mu$ from $\nu$ with respect to  $\gamma$. The greatest lower bound of the numbers $\Delta^\gamma(\mu,\nu)$ with respect to the set $\Gamma(\mu,\nu)$ of all possible couplings of $\mu$ and $\nu$ is then a natural distance between $\mu$ and $\nu$, known as the {\em Wasserstein 2-deviation}:
\begin{equation}\label{Wasserstein}
  \Delta(\mu,\nu)=\inf\{\Delta^\gamma(\mu,\nu)\,|\, \gamma\in\Gamma(\mu,\nu)\}.
\end{equation}
If $\nu=\delta_y$, then  $\gamma=\mu\times\delta_y$ is the only coupling of $\mu$ and $\nu$, in which case (\ref{Wasserstein}) reduces to (\ref{Dpoint}).

Strictly speaking, $\Delta(\mu,\nu)$ may fail to be a distance, since \eqref{Dpoint} can be infinite. But if one restricts $\Delta(\cdot,\cdot)$  to measures with finite standard deviations, then it becomes a proper metric \cite{Villani}. This metric also has the right scaling: if we denote the scaling of measures by $s_\lambda$, so that for $\lambda>0$ and measurable $X\subset\Rl$, $s_\lambda(\mu)(X)=\mu(\lambda^{-1}X)$, then $\Delta(s_\lambda\mu,s_\lambda\nu)=\lambda \Delta(\mu,\nu)$, showing that the metric is compatible with a change of units. Moreover, the metric is unchanged when both measures are shifted by the same translation.

If $\mu$ and $\nu$ have finite standard deviations, then the  Cauchy-Schwarz inequality gives the following bounds,
\begin{equation}\label{CSbounds}
\begin{split}
&(\Delta(\mu)-\Delta(\nu))^2+(\mu[x]-\nu[x])^2\leq\Delta(\mu,\nu)\\
&\quad\leq (\Delta(\mu)+\Delta(\nu))^2+(\mu[x]-\nu[x])^2,
\end{split}
\end{equation}
which are obtained exactly when there is a coupling giving perfect negative, resp. positive, correlation between the random variables in question, i.e., the  variables are linearly dependent.

\subsection{Errors as device figures of merit}

Given a distance  for probability distributions we can directly define a distance of observables $\sfe,\sff$,
\begin{equation}\label{distobs}
\Delta(\sfe,\sff)=\textstyle{\sup_\rho} \Delta(\sfe_\rho,\sff_\rho).
\end{equation}
 Note that we are taking here the {\it worst case} with respect to input states. Indeed, we consider the distance of an observable $\sff$ from an ``ideal''  reference observable $\sfe$ as a figure of merit for $\sff$, which a company might advertise: No matter what the input state, the distribution obtained by $\sff$ will be $\veps$-close to what you would get with $\sfe$. When closeness of distributions is measured by $\Delta(\cdot,\cdot)$, then \eqref{distobs} is the best $\veps$ for which this is true. As noted earlier, the distances $\Delta(\sfe_\rho,\sff_\rho)$ for individual states are practically useless as  benchmarks since the deficiencies of a device may not be detectable on a single state. However, these state-dependent measures may be useful if the goal is to control error or disturbance in a particular state. 

The additional maximization in \eqref{distobs} leads to some simplifications. Indeed, assume that $\sfe$ is a sharp observable  and that $\sff$ differs from $\sfe$ just by adding noise that is independent of the input state, that is, $\sff=\mu*\sfe$ for some probability measure $\mu$. Then  \cite{BLW2013b}
\begin{equation}\label{distobsEF}
  \Delta(\sfe,\mu\ast\sfe)=\Delta(\mu,\delta_0) =\sqrt{\mu[x^2]},
\end{equation}
so that $\Delta(\sfe,\mu\ast \sfe)\geq \Delta(\mu)$, and equality holds exactly in the unbiased case, $\mu[x]=0$.

\subsection{Calibration error}

The  supremum \eqref{distobs} over all states may not be  
easily accessible in experimental implementations. Therefore, it seems more reasonable to just calibrate the performance of a measurement of  $\sff$ as an approximate measurement of $\sfe$ by looking at the distributions $\sff_\rho$ for preparations for which $\sfe_\rho$ is nearly a point measure, i.e., those for which $\sfe$ ``has a sharp value''.\footnote{If  $\sfe_\rho$ is a point measure concentrated at $\xi$ then the effect $\sfe(\{\xi\})$ has eigenvalue 1 and $\rho$ is a corresponding eigenstate.} This can always be achieved when $\sfe$ is sharp, and in this case we are led to define the {\em calibration error} $\Delta_c(\sfe,\sff)$ of $\sff$ with respect to $\sfe$  as the greatest lower bound of   the $\veps$-calibration errors, $\veps>0$, as follows:
\begin{eqnarray}\label{calbratEps}
 \Delta^\veps(\sfe,\sff) &=&\sup\Bigl\{\Delta(\sff_\rho,\delta_y)\Bigm| y\in \Rl,\Delta(\sfe_\rho,\delta_y)\leq\veps\Bigr\}\quad \\
\Delta_c(\sfe,\sff)    &=& \inf\{\Delta^\veps(\sfe,\sff)\,|\, \veps >0\}\label{calbrate}
\end{eqnarray}
Provided that $\Delta(\sff_\rho,\delta_y)$ is finite for at least some $\veps>0$,  the limit in \eqref{calbrate} exists, because \eqref{calbratEps} is a monotonely decreasing function of $\veps$. Otherwise the calibration error is said to be infinitely large and $\sff$ is to be considered a bad approximation. In the finite case,
the triangle inequality gives that  $\Delta^\veps(\sfe,\sff)\leq\veps+\Delta(\sfe,\sff)$, and hence 
\begin{equation}\label{cal<Del}
  \Delta_c(\sfe,\sff)\leq \Delta(\sfe,\sff).
\end{equation}
From \eqref{distobsEF} we observe that if $\sff$ just adds independent noise to the results of $\sfe$, then $\Delta_c(\sfe,\sff)= \Delta(\sfe,\sff)$. In general, however, the inequality \eqref{cal<Del} is strict.

The Wasserstein distance of probability distributions may not at first sight be a practical quantity as it can be difficult to calculate directly. However, there is an alternative method of computing the error defined here as the infimum over all couplings; this is provided by Kantorovich's Duality Theorem \cite{Villani}, according to which this infimum over coupling measures is shown to be equal to the supremum over a certain set of functions. Illustrations  of this technique are found in our related works \cite{BLW2013b,BuPe14}.

\begin{example}\label{exa:indep-noise}\rm 
The method of adding independent noise  provides an important example of a joint approximate measurement of two observables. Consider any two sharp observables $\sfa$ and $\sfb$. If these observables do not commute in any state there is still the possibility that they can be measured jointly in an approximate way. In an approximate and unbiased von Neumann measurement of $\sfa$, with $U=e^{i\lambda A\otimes P_p}$, $Z=Q_p$, $\sigma=\kb\phi\phi$, the measured distribution is of the form $\mu*\sfa_\rho$; hence the measured observable is $\mu*\sfa$. Then we obtain $\Delta_c(\sfa,\mu*\sfa)=\Delta(\sfa,\mu*\sfa)=\sqrt{\mu[2]}$. The disturbance caused on $\sfb$ can be described in terms  of the distributions as $\sfb_\rho\mapsto \sfb_{\instru(\R)(\rho)}\equiv \sfb'_\rho$.

The observable $\sfb$ could also be measured approximately by an (unbiased) von Neumann measurement, realizing $\nu*\sfb$ as an approximation. It may happen that the measurements $\mu*\sfa$ and $\nu*\sfb$ can be combined into a joint measurement, in which case one has errors
\begin{align}\label{eqn:indep-noise}
\Delta(\sfa,\mu*\sfa)=\sqrt{\mu[2]},\quad \Delta(\sfb,\nu*\sfb)=\sqrt{\nu[2]}.
\end{align}
 For position and momentum this happens exactly when $\mu$ and $\nu$ are Fourier related \cite{CHT2005}, in which case
 $\sqrt{\mu[2]}\sqrt{\nu[2]}\ge \Delta(\mu)\Delta(\nu)\geq \frac \hbar{2}$. ---
\end{example}

\section{Comparison}\label{sec:critique}
 
 We now investigate the justification of the interpretation of $\epno$ as a putative state-specific quantification of measurement errors, and compare this quantity with the state-dependent distribution error based on the Wasserstein 2-deviation. Both quantities serve to define state-independent error indicators, which we will discuss later.
 
\subsection{Ways of expressing the noise-based error quantity}

We begin by writing the quantity $\epno$ in a variety of ways and proceed to interpret each of these forms. We introduce some shorthand notation:  $A_{\rm in}:=A\otimes\idty$, $A_{\rm out}:=U^*(\idty\otimes Z)U$, and $N(A):=A_{\rm out}-A_{\rm in}$ for the {\em noise operator}. Then we have, denoting by $\sfa=\sfe^A$ the sharp target observable and by $\sfc$ the approximating observable actually measured by the given scheme $\hM$:
\begin{align}
\hspace{-6pt}\epno(A,\hM,\rho)^2
&=\langle N(A)^2\rangle_{\rho\otimes\sigma}\label{eq:epno-def}\\
&=\int x^2\langle \sfe^{N(A)}(dx)\rangle_{\rho\otimes\sigma}\label{eq:noise-moment}\\
&=\int(x-y)^2\,{\rm Re}\langle \sfa_{\rm in}(dx)\sfa_{\rm out}(dy)\rangle_{\rho\otimes\sigma}\nonumber\\
&=\int(x-y)^2\,{\rm Re}\langle \sfa(dx)\sfc(dy)\rangle_\rho\label{eq:value-dev}\\
&=\langle A^2\rangle_\rho+\langle \sfc[x^2]\rangle_\rho- 2{\rm Re}\,\langle A \sfc[x]\rangle_\rho \label{eq:epno-A-C}\\
             &= \langle \sfc[x^2]-\sfc[x]^2\rangle_\rho  + \bigl\langle (\sfc[x]-A)^2\bigr\rangle_\rho. \label{OzaVeps1}
\end{align}
The first line is a compact rewriting of the definition of $\epno$ and the second gives this explicitly as the second moment of the distribution of the noise operator in the state $\rho\otimes\sigma$. In the next two lines we have introduced the bimeasure 
\begin{align}
(X,Y)\mapsto \xi^{\sfa,\sfc}_\rho(X,Y)&\equiv {\rm Re}\langle \sfa_{\rm in}(X)\sfa_{\rm out}(Y)\rangle_{\rho\otimes\sigma}\nonumber\\
&={\rm Re}\langle\sfa(X)\sfc(Y)\rangle_\rho\in[-1,1]\label{eq:bimeasure}
\end{align}
to write $\epno$ \textsl{formally} as a squared deviation  (which works mathematically since the integrand is separable). 
The last term of \eqref{eq:epno-A-C} arises from $\tr{(A\rho\otimes\sigma)U^*(\idty\otimes Z)U}$ and its complex conjugate by applying \eqref{obsScheme} with $A\rho$ replacing $\rho$. The last line expresses $\epno$ in terms of the intrinsic noise operator. This shows that $\epno$ depends only on the first two moment operators of $\sfa$ and $\sfc$. 

Essentially the only justification for the interpretation of $\epno$ as an error measure given by its proponents (e.g., \cite{Ozawa04}) is by making reference to the context of calibration  for the approximate measurement  of an observable $\sfa$. If the input state $\rho$ is an eigenstate of $A$, so that $\sfa_\rho$ is a point measure $\delta_a$, then one  has 
\begin{align}
\epno(A,\h M,\rho)^2&= \sfc_\rho[x^2]+a^2-2a\sfc_\rho[x]\nonumber\\
&= \int(x-a)^2\,\sfc_\rho(dx) =
\Delta(\sfc_{\rho},\delta_a)^2,\label{eps-o-class}
\end{align}
showing that $\epno$ corresponds to the classic Gaussian expression for the rms deviation from the ``true value''. In this special situation $\epno$ coincides thus with the Wasserstein 2-deviation $\Delta(\sfc_{\rho},\delta_a)$.  However,  in non-eigenstates, there is no ``true value''.

We note that similar expressions can be given for the noise-based disturbance quantity. We introduce the disturbance operator
$D(B):=B_{\rm out}-B_{\rm in}$, where $B_{\rm in}:=B\otimes\idty$ and $B_{\rm out}:=U^*B\otimes\idty U$. Denoting by $\sfb$ the
spectral measure $\sfe^B$ and by $\sfb'$ its distortion, $\sfb'=\stfm(\R)^*(\sfb(\cdot))$ by the instrument associated with $\hM$, we obtain:
\begin{align*}
\etno(B,\hM,\rho)^2
&=\int x^2\langle \sfe^{D(B)}(dx)\rangle_{\rho\otimes\sigma}\\
&=\int(x-y)^2\,{\rm Re}\langle \sfb_{\rm in}(dx)\sfb_{\rm out}(dy)\rangle_{\rho\otimes\sigma}\\
&=\int(x-y)^2\,{\rm Re}\langle \sfb(dx)\sfb'(dy)\rangle_\rho\\
&= \langle \sfb'[x^2]-\sfb'[x]^2\rangle_\rho  + \bigl\langle (\sfb'[x]-B)^2\bigr\rangle_\rho.
\end{align*}
Our subsequent discussion will focus mainly on $\epno$, with analogous comments applying to $\etno$.

\subsection{Limitations of the interpretation of the noise-based error}

The immediate quantum mechanical meaning of $\epno$ is that of being the square root of the {\em second moment} of the statistics obtained when the observable  associated with the (presumably) selfadjoint difference operator $N(A)=U^*(\idty\otimes Z)U - A\otimes\idty$ is measured on the system-probe state $\rho\otimes\sigma$. Hence, viewing the  definitions of $\epno,\etno$ from the perspective of classical statistical error analysis makes it extremely suggestive (perhaps almost irresistible) to consider them as ``natural'' quantum extensions of the notion of mean deviation between pairs of values of the input and output observables {\em measured jointly} on the same object -- hence as value deviations. 

However, as discussed in Subsection  \ref{sec:POVM}, one cannot, in general, assume the output operator $U^*(\idty\otimes Z)U$ and input operator $A\otimes \idty$ to commute, so that measuring the difference observable requires quite a different procedure  than measuring either of the two separate observables or than measuring them jointly (which is generally impossible). Neither of the three measurements will be compatible unless the output pointer and target observables do commute. It follows that the value of $\epno$ cannot be obtained from a comparison of the statistics of the  measurement $\hM$ and a control measurement of $\sfa$ in the state $\rho$. Put differently, declaring $\epno(A,\h M,\rho)$ to represent the error of $\hM$  as an approximate measurement of $\sfa$ in the state $\rho$ would be analogous to claiming that the measured values of the harmonic oscillator energy are equal to the sum of the values of the kinetic and potential energy (where these clearly have no simultaneous values). 

Thus, unless $\sfa$ and $\sfc$ are jointly measurable (at least in the particular state of interest), there is no justification to the claim that $\epno$ is a quantification of experimental error  -- notwithstanding the fact that this quantity can be experimentally determined itself. 

As similar discussion applies to the formulation of $\epno$ in terms of  \eqref{eq:bimeasure}. 
This bimeasure will not in general be a probability bimeasure as there will not be joint measurements of the respective pairs of observables $\sfa_{\rm in},\sfa_{\rm out}$ and $\sfa,\sfc$ unless they are compatible, which requires their commutativity. We note that the commutativity of $\sfa,\sfc$ is  related to that of $\sfa_{\rm in},\sfa_{\rm out}$ via the relation
\begin{align*}
&\langle \sfa_{\rm in}(X)    \sfa_{\rm out}(Y)  -  \sfa_{\rm out}(Y)   \sfa_{\rm in}(X)\rangle_{\rho\otimes\sigma}\\
&\qquad\qquad\qquad=\langle \sfa(X)\sfc(Y)-\sfc(Y)\sfa(X) \rangle_\rho.
\end{align*} 
Without the commutativity of $\sfa$ and $\sfc$, the terms appearing in \eqref{eq:epno-A-C} requires a measurement of the observable
given by $A\sfc[x]+\sfc[x]A$, which generally will not commute with either of the noncommuting operators $A$ and $\sfc[x]$; hence
the determination of $\epno$ via \eqref{eq:epno-A-C} is seen to require three incompatible measurements.

The unavailability of $\epno$ as a universally valid error measure may itself be construed as a quantum phenomenon. Consider a measurement of a sharp observable $C=\sfc[x]$ as an approximation of observable $\sfa$.  In that case $V(\sfc)=0$ and according to Eq. \eqref{OzaVeps1}  one has then $\epno(A,\hM,\rho)^2=\langle\psi|(A-C)^2\psi\rangle$ if $\rho$ is a pure state with associated unit vector $\psi$. For simplicity we assume that $A,C$ are bounded. The condition $\epno(A,\hM,\rho)=0$ implies that $A\psi=C\psi$, and if the spectral measures $\sfa,\sfc$ commute on $\psi$, this entails $A^n\psi=C^n\psi$ for all $n\in\mathbb N$, and this yields $\sfa_\rho=\sfc_\rho$. This is analogous to the classical case, where the vanishing of the squared deviation implies that the two random variables in question are equal with probability 1. Put differently, in classical probability, vanishing rms deviation of two random variables in a given probability distribution entails that the rms deviation between any functions of them vanishes as well. This is no longer true in quantum mechanics: if $A, C$ do not commute, then $\epno(A,\hM,\rho)=0$ only gives $A\psi=C\psi$ but generally $\sfa_\rho\ne \sfc_\rho$. We will give examples below showing that such false indications of perfect accuracy do happen.

In order to fix this deficiency, Ozawa \cite{ozawa2005perfect} has given a characterization of perfect accuracy measurements for a given pure state $\psi$ in terms of perfect correlations between input and output observable, in that state; he showed that these conditions can only be satisfied on states that are in the commutativity subspace of the two observable -- which therefore has to be nontrivial.\footnote{For an analysis of the commutativity subspace and the joint measurability of two sharp observables see \cite{Ylinen1985}.} Accurate measurements in such a state $\psi$ are then also characterized by the vanishing of $\epno$ on a suitable subspace of vectors in this commutativity subspace. This underlines the fact that $\epno$ is valid as an error measure  only to the extent to which the approximating observable commutes with the target observable.

\subsection{Ways of measuring noise-based error and disturbance}

a) {\em Directly measuring the noise operator.}
As noted above, the immediate meaning of $\epno$ is related to its expression as the expectation of the square of the noise operator, $\langle N(A)^2\rangle_{\rho\otimes\sigma}$. The experimental methods used by the Toronto group in confirming Ozawa's inequality \cite{Roz12} can be adapted to performing a direct measurement of $N(A)^2$.

b) {\em Method of weak values.} 
It was noted in \cite{LundWiseman2010} that the numbers $\xi^{\sfa,\sfc}_\rho(X,Y)\in[-1,1]$ can  be determined experimentally 
by application of weak measurements; then in the case of discrete finite observables, the integral (sum) form (\ref{eq:value-dev}) may be used to reconstruct the value of $\epno(A,\hM,\rho)$. This weak value method was first used in the experiment of \cite{Roz12}, in which $\epno,\etno$ are determined in this way.  However, in that case the approximators and target observables do actually commute, so that the numbers $\xi^{\sfa,\sfc}_\rho(X,Y)$ are in fact probabilities and could have been determined directly from sequential measurements instead.

c) {\em Three-state method.}
In response to comments on the interpretational problems associated with $\epno,\etno$ \cite{BuHeLa04,Werner04}, Ozawa \cite{Ozawa04} proposed a method of measuring $\epno$ that was later termed  \textsl{three-state method} by the experimenters who used it to measure $\epno$ and $\etno$ and test Ozawa's inequality \cite{Erh12}; it is encapsulated in the formula, obtained readily by further manipulation of \eqref{OzaVeps1}:
\begin{equation}\label{3-state}\begin{split}
\epno&(A,\h M,\rho)^2 =\tr{\rho A^2}+ \tr{\rho \sfc[x^2]}\\
&\qquad +\tr{\rho \sfc[x]}
+\tr{\rho_1 \sfc[x]} -\tr{\rho_2 \sfc[x]},
\end{split}
\end{equation}
where  the (non-normalized) states $\rho_1,\rho_2$ are given by $\rho_1=A\rho A$, $\rho_2=(A+\idty)\rho(A+\idty)$. While now the quantity $\epno$ is  manifestly determined by the statistics of $\sfa$ and $\sfc$, one can no longer claim it to be state-specific.
This is because now $\epno$ is a combination of numbers that are obtained from measurements performed on three distinct states $\rho,\rho_1,\rho_2$.

d) {\em Using sequential measurements.}
In the case of a discrete sharp target observable $\sfa$ (with complete family of spectral projections $A_i$) and commuting approximator $\sfc$, one can use a sequential measurement of $\sfa$ and then $\sfc$ to realize the joint (product) spectral measure 
defined by $X\times Y\mapsto \sfa(X)\sfc(Y)$ provided the first measurement is a L\"uders measurement, that is, its channel is $\rho\mapsto \sum_i A_i\rho A_i$. One can then  apply \eqref{eq:value-dev} to determine $\epno$.

Perhaps somewhat surprisingly, the same method can be used to obtain the disturbance measure $\etno(B,\hM,\rho)$ if the disturbed observable $\sfb'$ commutes with $\sfb$. This possibility was considered unavailable in \cite{LundWiseman2010} but shown to work in \cite{BuschStevens14} if $\sfb$ is sharp and discrete (with spectral projections $B_k$) and the L\"uders channel is used for the initial control measurement of $\sfb$. The task is to compare the values of  measurements of  $\sfb$ before and after a  measurement of $\sfc$ with instrument $\stfm$ (used to approximate $\sfa$). If $\sfb'=\stfm(\R)^*(\sfb(\cdot))$ commutes with $\sfb$, then the marginal joint observable for $\sfb$ and $\sfb'$ in this sequence of three measurements is in fact the product observable given by $X\otimes Y\mapsto \sfb(X)\sfb'(Y)$ and thus leads to a direct determination of $\etno$ as a value deviation measure.

\subsection{Commuting target and approximator}\label{sec:ozawa}

We now turn to the case of commuting target $\sfa$ and approximator $\sfc$. In this instance, as given above by the integral form \eqref{eq:value-dev}, $\epno$ has a probabilistic interpretation as value comparison error since $\xi^{\sfa,\sfc}_\rho$ extends to the quantum mechanical joint probability distribution of the two observables  $\sfa$ and $\sfc$. Since now $\xi^{\sfa,\sfc}_\rho$ constitutes a {\em coupling} $\gamma$ for $\sfa_\rho$ and $\sfc_\rho$ it follows that
\begin{equation}\label{Oz-commutative}\begin{split}
\epno(A,\hM,\rho)^2&= \sfa_\rho[x^2]+\sfc_\rho[x^2]-2\langle A\sfc[x]\rangle_\rho \\ 
& \Delta^{\gamma}(\sfa_\rho,\sfc_\rho) \geq \Delta(\sfa_\rho,\sfc_\rho).
\end{split}
\end{equation}
Thus, in this commutative case, the NO-error provides a simple upper bound for the state-dependent error $\Delta(\sfa_\rho,\sfc_\rho)$. This is in line with the fact that $\epno$ accounts for the correlation between $\sfa$ and $\sfc$ as well as preparation uncertainty, while $\Delta$ merely compares their distributions. Similar remarks apply to $\etno$. 

In the case of approximations with independent noise, represented by an approximator $\sfc=\mu*\sfa$ to a sharp target observable $\sfa$ (see Example \ref{exa:indep-noise}), one has
\begin{equation}\label{eq:epno-indep}
\epno(A,\hM,\rho)=\sqrt{\mu[2]}=\Delta(\sfa_\rho,\mu*\sfa_\rho).
\end{equation}
Observe that here the state-specific errors have become entirely state independent and the value comparison and distribution errors coincide.

\begin{example}\label{exa:Q vs -Q}\rm
It has been noted \cite{Roz13} that there are instances where the quantities $\epno,\etno$ are more sensitive to deviations between the  target and approximator observables than the Wasserstein 2-deviation. This is nicely illustrated with the following example, where the observable to be measured is position $Q$ and the approximator  is  the sharp observable $Q'=-Q$. Then for any state $\rho$ one has 
\begin{align*}
\epno(Q,\hM,\rho)^2&=\tr{\rho(Q-(-Q))^2}\\
&=4\sfq_\rho[x^2]=4\Delta(\sfq_\rho)^2+4{\langle Q\rangle_\rho}^2\,.
\end{align*}
Now, if the  density of  $\sfq_\rho$ is an even function, then $\epno(Q,\hM,\rho)=2\Delta(\sfq_\rho)$ while $\Delta(\sfq_\rho,\sfq^-_\rho)=0$ since the distributions coincide; here $\sfq^-$ is the spectral measure of $-Q$. Thus  $\epno$ is more capable of seeing the difference between $Q$ and $-Q$ in the present case, particularly in an even probability distribution. This is easily understandable since here the value comparison error analysis is available and provides more detailed information: the quantity $\epno$ captures  the strong anticorrelation between the jointly measured quantities $Q$ and $-Q$ that arises due to their functional dependence. By contrast, the quantity $\Delta(\sfq_\rho,\sfq^-_\rho)$ describes the deviation between the distributions $\sfq_\rho$ and $\sfq^-_\rho$, and thus vanishes if these distributions are even functions. ---

\end{example}

The following examples involve approximators and distorted observables that are trivial. These are of course very bad as approximations of sharp observables, but still this does not always show at the level of distributions. It will be seen that the value comparison method, which is applicable in these cases, is more sensitive in exhibiting the poor quality of trivial approximators. 
With both measures on can indeed verify that the approximations are trivial if one is allowed to test the devices on sufficiently many states.

\begin{example}\label{exa:trivial} {\rm Consider two sharp observables $\sfa$ and $\sfb$ and an arbitrary state, $\rho$. Define trivial observables $\sfc=\sfa_{\rho}\idty$, $\sfd=\sfb_{\rho}\idty$. Then, if the joint measurement $\hM$ of $\sfc$ and $\sfd$ is applied to the state $\rho$, the distributions of both $\sfa$ and $\sfb$ are accurately reproduced in that state. Hence there is no nontrivial bound to the combined distribution errors for two observables in an arbitrary state:  
$\Delta(\sfa_{\rho},\sfc_{\rho})=0=\Delta(\sfb_{\rho},\sfd_{\rho})$. By contrast, 
\begin{equation}\label{eqn:triv-approx}
\epno(A,\hM,\rho)^2=2\Delta(\sfa_\rho)^2;
\end{equation}
this quantity being nonzero reflects the independent contributions of the random spreads of $\sfa$ and $\sfc$ as they are being jointly measured. ---
}
\end{example}

Next we consider some model realizations of error- or disturbance-free joint measurements, while nevertheless the 
quantities $\epno$ and $\etno$ are nonzero in some or all states.

\begin{example}\label{exa:small-eps1}\rm
Here is an instance of a disturbance-free measurement where the measured observable is trivial; yet, for any given state the measurement can be adapted to reproduce the statistics accurately while the value comparison error $\epno\ne 0$.

Take the probe to be a system of the same kind as the object, $U$  the identity, $Z=A$.  This measurement scheme gives one and the same output distribution -- namely, $\sfa_{\sigma}$ -- for every input state $\rho$. Such a measurement is completely uninformative as it does not discriminate between any pair of different input states. In other words, the measured observable is trivial, $\sfc(X)=\sfa_{\sigma}(X)\,\idty$, and thus commutes with $A$. 

This model is comparable to a broken clock that works perfectly accurately every twelve hours -- except one cannot tell when this would be unless one knows the time by other means. Knowing that the  error is small for a set of input states with certain properties does not help unless one has the prior information that a given input state is from this class. 

The NO-error can be determined via the value-comparison method, that is, by measuring $\sfa$ on the object system jointly with measuring $\sfa$ on the probe. The value of $\epno(A,\hM,\rho)^2$ is
\begin{align*}
 \epno(A,\hM,\rho)^2&=  
\Delta(\sfa_{\rho})^2 + \Delta(\sfa_\sigma)^2+\bigl(\langle A\rangle_\rho-\langle A\rangle_\sigma\bigr)^2\\
&\ge \Delta(\sfa_\rho,\sfc_\rho)^2
= \Delta(\sfa_\rho,\sfa_\sigma)^2.
\end{align*}
This illustrates the different roles of the two state-dependent measures: $\Delta$ measures the difference between  the distributions $\sfa_\rho$ and $\sfc_\rho=\sfa_\sigma$, which are indicated as being identical when $\sigma$ is chosen to be equal to a given $\rho$. By contrast, $\epno$ shows that the two $\sfa$ measurements performed simultaneously on the object and probe are statistically independent giving three separate contributions to the measurement noise: the systematic error as the deviation between the mean values, the random noise arising from the probe preparation $\sigma$, and the preparation uncertainty of $\sfa$ arising from the state $\rho$ of the object.

Since the state does not get altered, one has $\sfb'=\sfb$ and so $\etno(B,\h M,\rho)=0$  and  $\Delta(\sfb_\rho,\sfb'_\rho)=0$ for any (sharp or unsharp) observable $\sfb$ in any state $\rho$. ---
\end{example}

We can now see how Ozawa's or Branciard's inequalities incorporate the possibility of vanishing disturbance (or error, as shown in the next example): when $\etno(B,\h M,\rho)=0$, the inequality reduces to 
\[
\epno(A,\hM,\rho)\Delta(\sfb_\rho)\ge \tfrac12\bigl|\langle[A,B]\rangle_\rho\bigr|\,;
\]
since $\epno$ carries a preparation uncertainty contribution, one has $\epno(A,\hM,\rho)\ge\Delta(\sfa_\rho)$, and the trade-off is seen to be one for preparation uncertainties rather than for error and disturbance. In fact, if $\sfa$ has an eigenvalue one can choose $\sigma$ to be an associated eigenstate, so that the random noise arising from the probe preparation vanishes, $\Delta(\sfa_\sigma)=0$; moreover for states $\rho$ with $\langle A\rangle_\rho=\langle A\rangle_\sigma$, then also the systematic error vanishes, and $\epno(A,\hM,\rho)$ is reduced to the pure preparation uncertainty $\Delta(\sfa_\rho)$.

\begin{example}\label{exa:small-eps2}\rm 
Next we construct an example of an accurate measurement which also has vanishing disturbance on a particular state while the NO-disturbance has a nonzero value.

Such a model is obtained by taking $U$ as the swap operation. Here we have $\sfc=\sfa$ and $\sfb'=\sfb_{\sigma}\idty$. This scheme gives a  NO-disturbance, $\etno(B,\hM,\rho)$, which is very small for some input states and a suitable probe state and becomes arbitrarily large on other states: 
\begin{align*}
 \etno(B,\hM,\rho)^2 &=\Delta(\sfb_\rho)^2+\Delta(\sfb_\sigma)^2+\bigl(\langle B\rangle_\rho-\langle B\rangle_\sigma\bigr)^2\\
&\ge  \Delta(\sfb_\rho,\sfb'_\rho)^2 = \Delta(\sfb_\rho,\sfb_\sigma)^2 .
\end{align*}
For $\sigma=\rho$,  $\Delta(\sfb_\rho,\sfb'_\rho)=0,$ indicating correctly that there is no disturbance in the distribution of $B$, 
while $\etno(B,\hM,\sigma )=
\sqrt{2}\Delta(\sfb_\sigma)$ indicates that the distorted observable $\sfb'$ has become statistically independent of the observable $\sfb$.  ---
\end{example}

Ozawa's inequality \eqref{eqn:Ozawa-inequality} has been presented as an invalidation of Heisenberg's error-disturbance relation. As we see in the present example, error and disturbance can easily be simultaneously small for particular choices of {\em individual} states, in particular, \textsl{small enough to violate any  Heisenberg type inequality of the form \eqref{eqn:false-ur}}. This is true for \textsl{any}  state-dependent measure of error and disturbance, including our measures $\Delta(\sfa_\rho, \sfc_\rho)$, see, for instance, \cite{Rud2013}.

The previous examples highlight the different purposes served by the state-dependent measures $\Delta$ and $\epno,\etno$. They also provide test cases showing that the Ozawa and Branciard inequalities do not universally represent ``pure'' error-error or error-disturbance trade-off relations but generally involve preparation uncertainties and may even sometimes reduce to the standard preparation uncertainty relation. 

In these examples, we have also seen that it is possible to isolate the systematic and random error parts from the preparation uncertainties contained in $\epno,\etno$; one may even have these genuine error contributions both vanish in suitable measurement schemes. This demonstrates that on individual states, perfectly error-free and disturbance-free measurements are in fact possible -- a result that goes beyond Ozawa's aim of showing that the error-disturbance product may vanish.

The fact that even a measurement of a trivial observable can mimic a perfectly accurate measurement in some states highlights the need to test a measuring device on a sufficiently rich variety of object states in order to be able to assess the accuracy and precision of the device. State dependent error measures can only answer rather more limited questions. In fact, the distribution deviation measure indicates merely how much the distribution of the approximating observable differs from that of the target observable. The value comparison error (where it can be applied) enables one to detect whether or not the approximating observable is correlated with the target observable; the method of its determination involves a joint measurement of the target $\sfa$ and the approximator $\sfc$; note that this also yields $\sfa_\rho$ and $\sfc_\rho$ and thus allow one to compute the distribution deviation.

\subsection{Unbiased approximator}\label{sec:unbiased}

The NO-error becomes more directly tied to the Wasserstein 2-deviation in the class of measurements with {\em constant bias}, characterized by the condition that $\sfc[x]-A$ is a constant, $c\idty$. Here one has:
\begin{equation}\label{eqn:epno-unbiased}
\epno(A,\h M,\rho)^2=\Delta(\sfc_\rho)^2-\Delta(\sfa_\rho)^2+c^2.
\end{equation}
In the unbiased case, $\epno^2$ coincides with the surplus variance of the approximator $\sfc$ over the target $\sfa$, a quantity that one could have considered independently as a distribution comparison error measure in this case.

The bounds for $\Delta(\sfa_\rho,\sfc_\rho)$ arising from \eqref{CSbounds} then also apply to $\epno(A,\h M,\rho)$; in fact, in the unbiased case $\epno$ is the geometric mean of these bounds and hence less flexible as an evaluation of the deviation than $\Delta$, but still gives a simple estimate of the latter. 

However, it is to be noted that the condition (a) of a good error measure is not met by $\epno$, even when this measure is restricted to unbiased approximators.
This will be demonstrated in Example \ref{exa:epno-unbiased} of Subsection \ref{sec:noncommut} below.
We therefore proceed to investigate further the true meaning of $\epno$ for unbiased approximators.

In the case of an unbiased approximator, $c=0$,  the expression \eqref{OzaVeps1} for $\epno$ reduces to $\epno(A,\hM,\rho)^2=\tr{\rho V(\sfc)}$. For unbiased joint approximations of two noncommuting observables the following result holds.

\begin{theorem}\label{thm:unbiased}
Let $\sfa,\sfb$ be sharp observables and $\sfg$ a joint observable for two unbiased approximations $\sfc,\sfd$ of $\sfa,\sfb$. Then the intrinsic noise operators of $\sfc,\sfd$ satisfy the trade-off
\begin{equation}\label{intnoise-ur}
\tr{\rho V(\sfc)}\tr{\rho V(\sfd)}\,\ge\,\tfrac14\left|\tr{\rho[A,B]}\right|^2.
\end{equation}
Furthermore, the standard deviations obey the uncertainty relation
\begin{equation}\label{full-ur}
\Delta(\sfc_\rho)\,\Delta(\sfd_\rho)\,\ge\,\left|\tr{\rho[A,B]}\right|.
\end{equation}
\end{theorem}
Versions of the inequalities \eqref{intnoise-ur} and \eqref{full-ur} have appeared for the special case of position and momentum in \cite{AK65} and with a rigorous proof in \cite{Stulpe1988};  proofs of different degrees of generality, rigor and elegance can be found in \cite{AG88,Ishikawa1991,Ozawa1991,Ozawa05,Hayashi2006,Polt2013}. 

Inequality \eqref{intnoise-ur} can be rewritten in terms of $\epno$ and $\etno$, thus confirming that \eqref{eqn:false-ur} holds in the unbiased case. We state this here for general joint measurements
with {\em unbiased approximators}:
\begin{equation}\label{eqn:unbiased-noise-ur}
\epno(A,\hM,\rho)\,\epno(B,\hM,\rho)\ge\tfrac 12\bigl|\tr{\rho[A,B]}\bigr|.
\end{equation}
However, we now see that this inequality is, in the first place, appropriately interpreted as a constraint on the intrinsic unsharpness of the approximators: one may say that if the approximators are emulating the targets too well -- here in the sense that the first moment operators coincide -- then the price arising from the noncommutativity of the target observables is that the approximators must be sufficiently unsharp. In the second place, when applied to unbiased approximators, $\epno$  gives an estimate of the {\em distribution comparison} error (see eq.~\eqref{eqn:epno-unbiased}), as it accounts for the intrinsic noise inherent in the approximating observable;  therefore inequality \eqref{eqn:unbiased-noise-ur} does also admit an interpretation as a joint-measurement error relation in the case of unbiased approximators. 

One would usually consider unbiasedness, or absence of systematic errors, to be a feature of a good approximate measurement. In that case, inequality \eqref{eqn:unbiased-noise-ur} constitutes a Heisenberg-type
measurement uncertainty relation, notably in the case of position and momentum. It seems puzzling that in order to obtain a violation of this inequality, one must search for joint approximate measurements where the quality of the approximators is degraded: systematic errors must be allowed! One explanation of this puzzle is apparent from the above examples: while for unbiased approximations the quantity $\epno$ comprises 100\%\ intrinsic noise and hence error, changing the approximators all the way to trivial ones transforms $\epno$ into a quantity that may contain 100\%\ preparation uncertainty; according to Ozawa's inequality this makes room for the other quantity to have vanishing error $\epno$; but in this case Ozawa's inequality has become an expression of preparation uncertainty.

\subsection{Noncommuting target and approximator}\label{sec:noncommut}

The limitations of $\epno$ and $\etno$ as state-specific measures of error and disturbance become manifest when target and measured observable do not commute, or similarly, where the disturbance is such that the distorted observable does not commute with the one prior to measurement. One can construct measurement schemes that are  evidently quite  bad approximations, leading to vastly different distributions, but nevertheless yield  small or even zero $\epno$ on some states. This can happen in a measurement  $\h M$ in which the measured approximator $\sfc$ is projection valued, so that the intrinsic noise term in \eqref{OzaVeps1} vanishes;
then $\epno(A,\hM,\rho)^2=\langle (\sfc[x]-A)^2\rangle_\rho$, and this vanishes when $\sfc[x]-A$ has eigenvalue zero and $\rho$ is an associated eigenstate. Note also that in such situations the value-comparison interpretation is not available as there are no jointly obtainable values.

\begin{example}\label{exa:small-eps3}\rm 

In this example the spectrum of the approximator observable $\sfc$  
is discrete while that of the target, position $\sfq$, is continuous. Hence  for every state $\rho$ the distributions $\sfq_\rho$ and $\sfc_\rho$ are vastly different but $\epno=0$ on some states.

Let $A=Q$, and assume $\sfc$ is the spectral measure of $Q'=Q+\alpha\bigl(P^2/2m+(m\omega^2/2)Q^2-(\hbar\omega/2)\idty\bigr)$, where $\alpha$ is a positive constant. Then $Q'-Q=\alpha\bigl(P^2/2m+(m\omega^2/2)Q^2-(\hbar\omega/2)\idty\bigr)$, and the square of this operator has vanishing expectation value for the ground state $\psi_0$ of the harmonic oscillator. Thus, $\epno(Q,\hM,\rho_0)=\langle (Q'-Q)^2\rangle_{\psi_0}=0$ for $\rho_0=\kb{\psi_0}{\psi_0}$. Having purely discrete spectrum, the sharp observable $\sfc$ is clearly a bad approximation to $\sfq$,  but the quantity $\epno$ does not notice this in the state $\psi_0$.---

\end{example}

The above failure depends on the noncommutativity of $Q'$ and $Q$.  Take again $A=Q$ and  $\sfc=\sfe^{Q'}$. If $Q'=f(Q)$, and $\epno(Q,\h M,\rho)=0$, then $f(x)=x$ almost everywhere with respect to $\sfq_\rho$, that is, $\sfc_\rho=\sfq_\rho$. For example, if  $Q'$ differs from $Q$ by a piecewise constant function of $Q$, defined as $Q'-Q=a\sfq(\R\setminus[-a,a])$, then in the interval $[-a,a]$, the measurements coincide but outside they differ by a constant value $a$. For all states $\rho=\kb\psi\psi$  given by functions $\psi$ that are localized in the interval $[-a,a]$ we have $\epno(Q,\h M,\rho)=0$ and $\sfc_\rho=\sfq_\rho$.

\begin{example}\label{exa:Q-Q'}\rm

In this example the approximate position measurement is sharp, and almost all states are measured with $\epno=0$. The pointer observable is the standard position observable, as is often assumed for ``pointers''.
Nevertheless the output distribution is different from the correct position distribution for every input state.

Consider a measurement interaction $U$ of the form $U=\swap (\idty\otimes V)$, where $\swap$ denotes the swap map, and since standard position $Q$ is measured on the pointer, we have  $U^*(\idty\otimes Q)U=(V^*QV)\otimes\idty$. This also holds for all functions of $Q$, so the resulting observable is sharp. The NO-error  is
\begin{equation}\label{exQQerr}
  \epno\bigl(Q,\hM,\psi\bigr)^2=\ip{\psi}{(Q-V^*QV)^2\psi}.
\end{equation}
It is possible to construct\footnote{The proof of this and the further claims made here involve some functional analysis and is deferred to the Appendix.} a unitary operator $V$ such that $V^*QV=Q+\kb\phi\phi$ for some (not necessarily normalized) nonzero vector $\phi$. Therefore, we will have $\epno=0$ for all input vectors $\psi$ orthogonal to $\phi$. For a suitable choice of $\phi$ the distributions of $Q$ and $Q'$ will be distinct for all input states.  ---

\end{example}

For an accurate measurement of some quantity $A$ Ozawa's inequality  reduces to
\[
\etno(B,\hM,\rho)\Delta(\sfa_\rho)\ge\tfrac12 \bigl|\langle [A,B]\rangle_\rho\bigr|.
\]
This was used in \cite{Ozawa03} to show, for a specific scheme realising an accurate position measurement, that one can have $\epno(Q,\h M,\rho)=0$ and arbitrarily small $\etno(P,\h M,\rho)$  by choosing states $\rho$ with sufficiently large standard deviation of the position. It is argued there that this phenomenon of a  disturbance-free, precise measurement may open up possibilities for  novel high resolution measurement methods. However, small and even vanishing values of $\etno(P,\h M,\rho)$ can still go along with significant disturbances, so that such a far-reaching conclusion seems unfounded. Examples for this can be constructed in analogy to Examples \ref{exa:small-eps3} and \ref{exa:Q-Q'}.

Our next example shows that even within the restricted class  of unbiased approximators, $\epno$ may wrongly indicate perfect accuracy.

\begin{example}\label{exa:epno-unbiased}\rm

In this example we construct an approximator observable $\sfc$ that is unbiased with respect to the observable $\sfa$ defined as the spectral measure of $A:=\sfc[x]$ but nevertheless does not commute with $\sfa$ and yet $\epno(A,\hM,\rho_0)^2=0$ for some state $\rho_0$.

Let $\hi=\mathbb{C}^2$ and define $\sfc$ as the three-outcome observable 
\begin{align*}
1\mapsto C_1&=\gamma\tfrac12(\idty+\sigma_1),\\
 -1\mapsto C_2&=\gamma\tfrac12(\idty+\sigma_2),\\
 0\mapsto C_3&=2(1-\gamma)\tfrac12\bigl(\idty-\tfrac1{\sqrt2}(\sigma_1+\sigma_2)\bigr),\\
 \text {where }\ \gamma=2-\sqrt2.
\end{align*}
Here $\sigma_1,\sigma_2$ denote the first two Pauli matrices. Noting that $\gamma=\sqrt2(1-\gamma)$ one confirms immediately that $C_1+C_2+C_3=\idty$. Note that $C_1,C_2,C_3$ are positive rank-1 operators. Next we compute:
\begin{align*}
\sfc[x]&=\tfrac12\gamma(\sigma_1-\sigma_2)=:A,\\
\sfc[x]^2&=\tfrac12\gamma^2\idty,\\
\sfc[x^2]&=\gamma\bigl(\idty+\tfrac12(\sigma_1+\sigma_2)\bigr).
\end{align*}
It follows that 
\[
\sfc[x^2]-\sfc[x]^2=2(1-\gamma)\tfrac12\bigl(\idty+\tfrac1{\sqrt2}(\sigma_1+\sigma_2)\bigr).
\]
This is a rank-1 positive operator and the eigenstate associated with the eigenvalue zero is given by the projector
\[
\rho_0=\tfrac12\bigl(\idty-\tfrac1{\sqrt2}(\sigma_1+\sigma_2)\bigr).
\]
Therefore, $\epno(A,\hM,\rho_0)=0$, despite the fact that $\sfc$ is an obviously bad approximator to $\sfa$ and the distributions $\sfa_{\rho_0}$ and $\sfc_{\rho_0}$ are different. ---
\end{example}

Finally we show how an experimenter could achieve joint approximations with both NO-errors vanishing while these approximations are actually quite bad.

\begin{example}\rm
In Example \ref{exa:small-eps3}, observable $\sfc$ was defined as the spectral measure of $\sfc[x]=Q'=Q+\alpha\bigl(P^2/2m+(m\omega^2/2)Q^2-(\hbar\omega/2)\idty\bigr)$, and this was used to approximate $A=Q$. One may take the same observable $\sfc$  to also approximate $B=Q+\beta(P^2/2m+(m\omega^2/2)Q^2-(\hbar\omega/2)\idty\bigr)$, which for $\beta>0$  is again a shifted and scaled harmonic oscillator Hamiltonian.
If $\beta\ne\alpha$, we have that the difference operator $C[x]-B=(\alpha-\beta)\bigl(P^2/2m+(m\omega^2/2)Q^2-(\hbar\omega/2)\idty\bigr)$, so that for the ground state $\psi_0$ of the standard harmonic oscillator we have $\epno(A,\hM,\psi_0)^2=\langle(\sfc[x]-Q)^2\rangle_{\psi_0}=0$ and also $\epno(B,\hM,\psi_0)^2=\langle(\sfc[x]-B)^2\rangle_{\psi_0}=0$. Yet again the distributions of $B$
and $\sfc$ in the state $\psi_0$ are quite different. ---
\end{example}

The vanishing of both NO-errors in this example suggests perfect accuracy, and both the Ozawa and Branciard inequalities become tight, assuming value zero on both sides. Given that the approximations in the state $\psi_0$ are anything but good, one must conclude that these inequalities are not always meaningful as error trade-offs, even at their tight limits.

However exotic or artificial one may consider the measurement schemes constructed above to be, they constitute theoretical possibilities and thus test cases against which the suitability of any putative measure of error and disturbance could and should be considered. The above examples show that the quantities $\epno,\etno$ are unsuitable as universal benchmarks for error and disturbance of a measurement scheme $\hM$, particularly in a single state; they may vanish in cases where the measurements are clearly not accurate. The final example highlights a limitation of the scope of the Ozawa and Branciard inequalities as meaningful error trade-offs.

\subsection{Noise-based errors in qubit experiments}\label{sec:qubit}
The values of $\epno$ and $\etno$ have been determined for qubit measurements,  using the three-state method in an experiment carried out in Vienna \cite{Erh12,Vienna2} and the weak measurement method in Toronto \cite{Roz12}.  

The experiments are realizations of spin-1/2 and polarization observables, respectively. A detailed analysis
was carried out in \cite{BLW2014}. In the Vienna experiment a projective (von Neumann-L\"uders) measurement of a sharp observable $\sfc$ is performed as an approximation of a sharp observable $\sfa$ on the states $\rho,\rho_1,\rho_2$, as described in eq.~\eqref{3-state}; the required moments of $\sfc$ and $\sfa$ are obtained from the statistics of this measurement and a direct accurate $\sfa$ measurement. Similarly one obtains the moments of $\sfb,\sfb'$ by measuring the observable $\sfb$ on the required states directly and after the $\sfc$ measurement.

The approximate measurement investigated in the Toronto experiment is found to constitute an approximate joint measurement of sharp qubit observables $\sfa,\sfb$ (with values $\pm 1$) by means of compatible observables $\sfc,\sfd$. Here the pairs $\sfa,\sfc$ and $\sfb,\sfd$ do actually commute, so that the value comparison method based on sequential measurements would be applicable. However, the experimenters chose to use the indirect method of weak values to determine the values of $\epno(A,\hM,\rho)$ and $\etno(B,\hM,\rho)$.

It is instructive to compare the NO-errors with the Wasserstein deviations for these experiments.

We use Bloch sphere notation to write the spectral projections of $A=\boa\cdot\bosig$ as $A_\pm=\frac12(\idty\pm\inpr\boa\bosig)$, 
so that $A=A_+-A_-$,
and similarly for an observable $B=\bob\cdot\bosig$, where $\boa,\bob$ are unit vectors. For optimal approximations the approximators $\sfc,\sfd$ need to be assigned the same values $\pm 1$, so that, for example, $\sfc$ is given as a map $\pm 1\mapsto C_\pm$, with the positive operators $C_+=\frac12(c_0\idty+\inpr\boc\bosig)$, $C_-=\idty-C_+$.  (Positivity of $C_\pm$ is equivalent to  $\no{\boc}\le\min\{c_0,2-c_0 \}\le1$.) The Wasserstein deviation between $\sfa_\rho$ and $\sfc_\rho$ for a state $\rho=\frac12(\idty+\bor\cdot\bosig)$ is then
\[
\Delta\bigl(\sfa_\rho,\sfc_\rho\bigr)^2=2|1-c_0+\bor\cdot(\boa-\boc)|,
\]
which gives
\[
\Delta\bigl(\sfa,\sfc\bigr)^2=2|1-c_0|+2\no{\boa-\boc}.
\]
The best joint approximations are obtained for covariant approximators (see \cite{BLW2014}), where a covariant observable $\sfc$ is characterized by $c_0=1$.
The experiments quoted are using such approximators. The quantity $\epno(A,\h M,\rho)$ is then readily computed using \eqref{OzaVeps1}:
\begin{align}
\epno(A,\hM,\rho)^2
&=1-\no{\boc}^2\ +\ \no{\boa-\boc}^2\nonumber\\
&=\langle V(\sfc)\rangle_\rho+\tfrac1{4}\Delta(\sfa,\sfc)^4,\label{eq:epno-delta-qubit}\\
\langle V(\sfc)\rangle_\rho&=1-\no{\boc}^2.\nonumber
\end{align}
Here we see that $\epno$   is  in fact {\em state-independent!} This quantity is a mix of an error contribution and the intrinsic noise of the approximator observable -- which is already accounted for in the $\Delta$ term; it is not hard to see that $\epno(A,\hM,\rho)\le \Delta(\sfa,\sfc)$. 
For approximators that are smearings of the target observable, so that $\boc=\gamma \boa$, one has in fact $\epno(A,\hM,\rho)=\Delta(\sfa,\sfc)$. This situation arises in the Toronto experiment. 

We thus see that in the particular case of covariant qubit observables, $\epno$ has lost what the advocates of this measure consider to be one of its virtues: its state-dependence. It was already manifest in light of the availability of the three-state method that $\epno$ cannot be expected to be sensitive to differences in the observables being compared on a particular state. In fact, $\epno$ cannot capture the peculiar situation that was noted to arise in both the Vienna and Toronto experiments, where the input and output distributions are identical, so that the state-dependent (distribution) error vanishes. 

From the perspective of someone interested in assessing the overall performance of a measuring device, this apparent deficiency of $\epno$ turns out to be an advantage: instead of having to probe the whole state space, one can just apply the three-state method to obtain the worst-case error.

In both the Vienna and Toronto experiments \cite{Erh12,Roz12}, the quantities $\epno,\etno$ are carefully determined. However, the experimenters do not report any attempt to confront these values  with an actual estimation of errors for the measured observables $\sfc,\sfd$ as  approximations to the target observables $\sfa,\sfb$. Without such a comparison, a test of Heisenberg-type error-disturbance relations is not complete.

\section{Quantum measurement uncertainty}\label{sec:theorems}

We next present some theorems highlighting general aspects of the measurement uncertainty theme.  We will focus on the disputed error-disturbance relations and the closely related approximate joint measurement problem. By comparison, the preparation uncertainty relation is uncontroversial. The Kennard-Weyl-Robertson relations have been firmly established  as rigorous consequences of the quantum formalism. We only stress that the idea of preparation uncertainty is not exhaustively formalized in these relations either, and further aspects are elucidated in  alternative forms, such as entropic uncertainty relations \cite{Hirschman,Bialynicki2011,KrishnaPartha2002,Berta-etal2010} or trade-off relations for the overall widths of the distributions concerned \cite{LP1961,Cowling,UH1985}. An excellent review of such relations is given in \cite{Folland}.

\subsection{Structural measurement limitations}

Heisenberg's considerations concerning measurement uncertainty can be cast  readily in the operational language of quantum mechanics. His basic observation, namely that good measurements necessarily disturb the system, holds as a general principle, not just for position and momentum. It is expressed in the slogan ``{\em No measurement without disturbance\/}'', stated precisely as follows: if  the measurement is disturbance free, in the sense that $\instru(\Omega)(\rho)=\rho$ for all input states $\rho$, then the measured observable is trivial, that is, $\sff(X)=\mu(X)\idty$, for some probability measure $\mu$. Put differently, if a measurement tells us anything at all about the input, in the sense that the distribution of outcomes depends in some way on the input state, then some states must be changed through the measurement.\footnote{The authors agree that this was well but perhaps not widely known in the mid 1990s, if not earlier. An explicit statement with proof sketch appears in the 1996 second edition of \cite{QTM}.}

The above ``folk theorem''  would be practically worthless, however, if it were restricted to the completely disturbance-free case. Fortunately, it can be extended to the statement  that ``small disturbance implies small information gain''. One straightforward formulation runs as follows. The disturbance will basically be the largest change of output versus input state, measured in trace norm, and allowing input states to be entangled with some other system. That is, the disturbance of a channel $T$, a trace preserving completely positive map on the trace class, is set to be 
\begin{equation}\label{cbnorm}
  \cbnorm{T-\id}=\sup\tr{(T\otimes\id_n(\rho)-\rho)B}, 
\end{equation}
where the supremum runs over all integers $n$, density operators on $\hi\otimes\Cx^n$ and operators $B$ on that space with $||B||\leq1$. The index stands for ``complete boundedness'' \cite{Paulsen}, and the same expression has also been introduced by Kitaev as the ``diamond norm''. For the output probability measures we use the total variation norm $||\cdot||_1$. Then we have the following Theorem \cite{KSW08a}, which is proved, not surprisingly, by establishing a continuity property for the Stinespring dilation. 

\begin{theorem}\label{thm:smalldisturb}
Let $\instru$ be an instrument with the property that $\cbnorm{\instru(\Omega)-\id}\leq\veps$. Then there is a probability measure $\mu$ on the outcome space $\Omega$ such that, for every input state $\rho$ one has $||\sff_\rho-\mu||_1\leq\sqrt{\veps}$, where $\sff$ is the observable defined by $\instru$. 
\end{theorem}

In some sense this is a universal measuring uncertainty relation. It shows that there is some truth in Heisenberg's paper regarding the disturbance by measurement. However, it demands much more of a low disturbance measurement than just ``low disturbance of momentum'', and in return gives a much stronger result than ``poor measurement of position''. Therefore, more specific results, especially aimed at the position and momentum pair, will be given below (Sect.\ref{QPUR}).  

Given the maximality of the position and momentum observables $Q,P$, Proposition \ref{prop:productform} has a  dramatic consequence for their sequential measurements.

\begin{proposition}\label{prop:Q-P-maxdisturb}
Let  $\hM$ be a measurement scheme realizing an accurate measurement of position $Q$, with instrument $\instru$.  Then  for any observable $\sfg$ measured  after the execution of $\hM$,  the effects $\sfg'(Y) =\instru(\R)^*(\sfg(Y))$  of the distorted observable $\sfg'$ are functions of $Q$. 
\end{proposition}
Thus, whatever $\sfg$ is chosen for the second measurement, $\sfg'$ is a poor approximation of $P$. This measurement $\hM$ completely obviates the  momentum distribution associated with input state $\rho$. Similarly, any accurate momentum measurement destroys all the information about the position distribution of the input state $\rho$.

\subsection{Covariant phase space observable}

The prime example, for the purpose of this paper, of a joint observable is that of a covariant phase space observable, which represents a joint measurement of some {\em smeared}, or {\em fuzzy} versions of position and momentum. We recall briefly the definition and a characterization of such observables \cite{Davies,Holevo,QHA,Cassinelli2003,KLY2006}.

By a covariant phase space measurement we mean a measurement applicable to a quantum particle that has a characteristic transformation behaviour under translations of both position and momentum. Thus, if the measurement is applied to an input state shifted in position by $\delta q$ and in momentum by $\delta p$, the output distribution will look the same as without the shift, except that it is translated by $(q,p)\mapsto(q+\delta q,p+\delta p)$. This symmetry is implemented by the unitary Weyl operators (or Glauber translations) 
\[
W(q,p)=e^{{\frac i\hbar}{\frac{qp}2}}e^{-\frac{i}{\hbar}qP}e^{\frac{i}{\hbar}pQ}
\] 
acting in the $L^2(\Rl)$-representation of the particle's Hilbert space as
\[
(W(q,p)\psi)(x)=e^{-\frac{i}{\hbar}(\frac{qp}2-px)}\, \psi(x-q).
\]
Then the whole observable can be reconstructed from its operator density at the origin \cite{Holevo,QHA}, which must be  a positive operator $\tau$ of trace $1$ (i.e., a density operator as for a quantum state), up to a factor of $(2\pi\hbar)^{-1}$. The probability for outcomes in a set $Z\subseteq\Rl^2$ is then given by the  positive operator
\begin{equation}\label{povm}
  \sfm^\tau(Z)=\frac{1}{2\pi\hbar}\int_Z\mskip-5mu   
  W(q,p)^*\tau W(q,p) \,dqdp . 
\end{equation}
The  property that allows the interpretation of such measurements as approximate joint position-momentum measurements is the form of their marginals $\sfm_1^\tau$, $\sfm_2^\tau$, which are convolutions of the form $\sfm_{1,\rho}^\tau=\mu_\tau*\sfq_\rho$ and $\sfm_{2,\rho}^\tau=\nu_\tau*\sfp_\rho$, with  $\mu_\tau=\sfq_{\Pi\tau\Pi^*}$ and $\nu_\tau=\sfp_{\Pi\tau\Pi^*}$, where $\Pi$ is the parity operator, $(\Pi\psi)(x)=\psi(-x)$. As a consequence of eqs.~\eqref{eqn:indep-noise} and \eqref{eq:epno-indep}, we then
have the same Heisenberg-type inequality for both Wasserstein deviations and NO-errors, which here are state-independent:
\begin{equation}\label{eqn:phase-space-ur}
\mu_\tau[x^2]\,\nu_\tau[x^2] \ge(\Delta\mu_\tau)^2\,(\Delta\nu_\tau)^2
\ge\frac{\hbar^2}4.
\end{equation} 
The first inequality becomes an equation if the measurements are unbiased, $\mu_\tau[x]=\nu_\tau[x]=0$. If in addition $\tau$ is the ground state of the harmonic oscillator, then also the second inequality becomes an equation.  In this case, the associated phase space distribution  is named after Husimi, who discovered it in 1940 \cite{Husimi}.

We note that any covariant phase space observable $\sfm^\tau$ can be implemented as the high amplitude limit of the signal observable measured by an eight-port homodyne detector \cite{Caves1994}; for a rigorous proof of this statement, see \cite{KL2008b}. 
Another model realization of covariant phase space observables is provided by the Arthurs-Kelly model \cite{AK65}. This was shown in \cite{Busch82} in the case where the initial state of the two probes is a pure product state and in \cite{BuBu2014} for arbitrary probe states.

\subsection{Joint measurement relations}\label{QPUR}

The following measurement uncertainty relations for $Q$ and $P$ were proven for state-independent calibration errors and the maximized Wasserstein 2-deviations in \cite{BLW2013b,BLW2013c}.

\begin{theorem}[Measurement Error Relations] 
Let $\sfm$ be any observable with outcome space $\R^2$. Then
\begin{equation}\label{MeasURc}
  \Delta_c(\sfq,\sfm_1)\Delta_c(\sfp,\sfm_2)\geq \frac\hbar2
\end{equation}
and 
\begin{equation}\label{MeasUR}
  \Delta(\sfq,\sfm_1)\Delta(\sfp,\sfm_2)\geq \frac\hbar2
\end{equation}
whenever the terms on the left hand sides are finite.
In both cases equality holds for a covariant phase space measurement $\sfm^\tau$ whose generating density $\tau$ is the ground state of the operator  $H=Q^2+P^2$.
\end{theorem}

It is not hard to see that if one of the error terms  tends to zero, that is, the corresponding marginal is (nearly)  error-free, then the other error becomes  infinite in the limit, so that the above inequalities hold in these limiting cases. This can be shown explicitly using Proposition \ref{prop:Q-P-maxdisturb} when one or the other error is actually zero; in this case the other error is infinite.

It must be noted that the above joint measurement trade-off relation for maximized 2-deviations is an idealization: for  realistic measurements $\sfm$  with finite operating ranges, the quantities appearing in eqs.~\eqref{MeasURc} and \eqref{MeasUR} will be infinite, so that these inequalities become trivial. 
In the general case of a phase space measurement with finite operating range, the task of proving a nontrivial error trade-off relation can be approached by restricting the supremum of the Wasserstein 2-deviations to those states that are localized within the operating range, thus yielding finite errors. While a proof of  measurement uncertainty relations for Wasserstein 2-deviations amended along these lines is presently outstanding, we expect it will work in a similar way to the approach taken by Appleby in the case of the maximized NO-errors; we shall briefly review this next.

Appleby \cite{Appleby1998b} gives a proof sketch for the trade-off relation
\[
\epno(Q,\hM)\,\epno(P,\hM)\ge \frac \hbar2
\]
for any approximate joint measurement $\hM$ of position and momentum, where $\epno(Q,\hM)=\sup_\rho\epno(Q,\hM,\rho)$,  $\epno(P,\hM)=\sup_\rho\epno(P,\hM,\rho)$. He then proceeds to indicate how similar arguments can be used to obtain a trade-off for measurements with finite ranges, characterized by the restriction of states $\rho$ to those whose first moments $\langle Q\rangle_\rho$, $\langle P\rangle_\rho$ are bounded within fixed intervals of sizes $\delta q$ and $\delta p$ and whose variances $\Delta(\sfq_\rho)$ and $\Delta(\sfp_\rho)$ are not greater than given numbers $\Delta q$ and $\Delta p$, respectively (where $\Delta q\Delta p\ge\hbar/2$):
\[
\left(\epno'(Q,\hM)+\frac\hbar{\delta p}\right)\left(\epno'(P,\hM)+\frac\hbar{\delta q}\right)\ge\frac\hbar 2\left(1+\frac{2\hbar}{\delta q\delta p}\right).
\]
Here $\epno'(Q,\hM)$, $\epno'(P,\hM)$ are the suprema over all states that satisfy the above constraints. It is clear that in the limit $\delta q\to\infty$, $\delta p\to\infty$, the previous idealized inequality is recovered.

It is a curiosity that the maximized NO-error is a reliable indicator of the presence or absence of differences between the target and approximator observables, despite the fact that the error interpretation of the state dependent quantities used for its determination is not generally applicable. It is worth noting here that the example of Ozawa's and Branciard's inequalities highlights the advantages of joint measurement trade-off relations for state-specific errors as the latter typically do have finite values for a large class of states (this holds notwithstanding the provisos we have pointed out regarding these specific relations).

Error trade-off relations have also been proven for approximate joint measurements  of a pair of $\pm 1$-valued qubit observables $\sfa,\sfb$ \cite{BuHe08,BLW2014}. In such a case the product of deviations does not possess  a nontrivial bound, so that it is more informative to minimize the \textsl{sum} of the (squared) errors. This yields the following result for the qubit observables $\sfa,\sfb$, with the notations of Sect.\ref{sec:qubit}.
\begin{theorem}[Qubit Error Relation]\label{thm:qubit-error}
Let $\sfm$ be any approximate joint measurement of the $\pm1$-valued qubit observables $\sfa,\sfb$. Then 
\begin{align*}
\Delta\bigr(\sfa,\sfm_1\bigr)^2&+\Delta\bigr(\sfb,\sfm_2\bigr)^2\ \ge\ \sqrt{2}\bigl[\no{\boa-\bob}+\no{\boa+\bob}-2  \bigr].    
\end{align*}
This bound is tight and quantifies the degree of incompatibility of $\sfa,\sfb$. It can be satisfied when the approximators $\sfm_1,\sfm_2$ are covariant.
\end{theorem}

As a consequence of Theorem \ref{thm:qubit-error}  and the close quantitative connection between $\epno$ and the $\Delta$ distance given in eq.~\eqref{eq:epno-delta-qubit}, it turns out that the NO-errors  obey a Heisenberg-type trade-off themselves: in any joint approximate measurement $\hM$ of two qubit observables $\sfa,\sfb$, with covariant approximators one has \cite{BLW2014}
\begin{align*}
\epno(A,\hM,\rho)&+\epno(B,\hM,\rho)\nonumber\\
& \ge\ \frac 1{\sqrt{2}}\bigl[\no{\boa-\bob}+\no{\boa+\bob}-2  \bigr].
\end{align*}
This is an inequality in the spirit of Heisenberg's original ideas, in that it states a trade-off between the approximation errors in an approximate joint measurement of incompatible observables $\sfa,\sfb$, where the bound is determined by their degree of incompatibility. Given that Branciard's tight inequality for the case of qubits is compatible with our Heisenberg-type relation, it must be seen as a confirmation rather than a violation of Heisenberg's ideas.

\section{Conclusion}\label{sec:final}

We have investigated what is required for establishing Heisenberg-type error-disturbance  relations as rigorous consequences of quantum mechanics, and have reviewed forms such relations on the basis of  two proposed quantum generalization of Gauss' classic root-mean-square deviation. 

We have compared  definitions of measurement error and disturbance in terms of Wasserstein 2-deviations with the definitions based on the expectations of the squared noise and disturbance operators.
In both cases, state-dependent and state-independent versions are available, the latter being defined as maxima over all states of the
respective state-dependent quantities. The Wasserstein 2-deviation is conceived as a distribution comparison measure that can be applied to all approximators of a given observable. The noise-based quantities $\epno$, $\etno$ are best understood as value comparison measures, and as such they are only applicable in cases where the target and approximator observables are compatible. Within this constraint, value comparison can be more informative than mere distribution comparison as its method employs joint measurements on the same system rather than separate measurements performed on distinct systems in the same state. 

Even where the value-comparison method is applicable, $\epno$ and $\etno$ are not always purely measures of error and disturbance alone since they also contain preparation uncertainty contributions.  It follows that Ozawa's and Branciard's inequalities do not represent a pure form of error trade-off for joint approximate measurements, particularly, due to the presence of preparation uncertainties besides the error contributions. 

We have shown that $\epno,\etno$ become unreliable as indicators of error and disturbance in the case of noncommuting target and approximator observables; this entails that the Ozawa and Branciard inequalities cannot claim universal validity.

We take the limitation  of the applicability of $\epno$, $\etno$ as error and disturbance measures as a demonstration of  the limitations of the {\em observable-as-operator} point of view that has so long dominated the teaching of quantum mechanics. 
However, it was also noted in the last section that these limitations  do not apply to the maximized noise-based error measure, as proposed and developed by Appleby \cite{Appleby1998a,Appleby1998b}. Universal joint measurement uncertainty relations have been established for maximized NO-errors and maximized Wasserstein deviations in the case of  position and momentum and of qubit observables.

Since the noise-operator based measures have until recently been the only candidates considered as state-dependent value comparison errors, the question whether quantum mechanics entails nontrivial error and disturbance bounds for joint measurements on individual states must be considered an open problem. As we have seen, the distribution comparison error measures cannot be expected to obey nontrivial, unconditional uncertainty relations.

We note that since the publication of the preprint for our \cite{BLW2013c}, there has been growing critical awareness of the shortcomings of the quantities $\epno,\etno$, with some similar comments and analyses as given here (e.g., \cite{Dressel2014,Rud2013}). Our analysis is a development of arguments that were presented in \cite{BuHeLa04,Werner04}, which were largely misunderstood in that our criticism of what we referred to as lacking operational significance (the failure of $\epno$ to reliably indicate the presence or absence of errors for all approximators) was wrongly taken as an assertion that the quantities $\epno,\etno$ were not accessible to experimental determination.

As interesting venues for further research into uncertainty relations we mention possibilities of defining measures of error and disturbance other than those based on the Wasserstein 2-deviation.  Very recently, trade-off relations in the spirit of our calibration relation were formulated and proven for \textsl{entropic} measures of error and disturbance \cite{Buscemi2014,Coles2013}. In \cite{Ipsen2013}, the total variation norm is used to deduce error trade-off relations for discrete observables for finite dimensional systems. The concept of \textsl{error bar width}, introduced in \cite{BuPe07} to formulate a calibration error relation for position and momentum, was adapted to yield generic joint measurement error relations for arbitrary pairs of discrete observables in finite dimensional Hilbert spaces in \cite{Miyadera2011}. Yet another recent line of research has led to uncertainty relations in the context of quantum estimation theory (e.g., \cite{hofmann2003uncertainty,watanabe2011uncertainty,Dressel2014}), which concern parameter comparisons in contrast to comparisons of observables that are the focus of the present study.

To conclude, there remain many interesting open questions concerning quantum measurement uncertainty, not least the problem of casting and rigorously proving error and disturbance relations  for measurements with finite operating ranges.

\acknowledgements{We are indebted to four anonymous referees for their valuable criticisms and perceptive observations, which have led to numerous improvements and clarifications. Thanks are also due to Johannes Biniok, Thomas Bullock and Roger Colbeck for their careful scrutiny of the near-final manuscript. This work was partly supported (P.B., P.L.) by the Academy of Finland, project no 138135, and EU COST Action MP1006.
R.F.W.\ acknowledges support from the European network SIQS. This work has benefited from discussions at a conference organised as part of COST Action MP1006, {\em Fundamental Problems in Quantum Physics}. }

\section*{Appendix: Proof construction for Example \ref{exa:Q-Q'}}

There are two claims made in the text of Example \ref{exa:Q-Q'}.
\begin{itemize}
\item[(A)]For suitable $\phi$ the operators $Q$ and $Q'=Q+\kb\phi\phi$ are unitarily equivalent, so that $Q'=V^*QV$ for some unitary operator $V$.
\item[(B)]The probability distributions for $Q$ and $Q'$ are different for all input pure states, provided $\phi$ is suitably chosen.
\end{itemize}

For both issues it will be helpful to consider the resolvents of $Q$ and $Q'$.
For $z\in\Cx\setminus\Rl$ we denote the resolvents by $R_z=(Q-z\idty)^{-1}$ and $R'_z=(Q'-z\idty)^{-1}$. Then
\begin{eqnarray}\label{Krein}
  R'_z-R_z&=-R'_z\kb\phi\phi R_z \nonumber\\
  R'_z\phi&=\Bigl(1+\ip{\phi}{R_z\,\phi}\Bigr)^{-1}\,R_z\phi \nonumber\\
  R'_z  &= R_z- \frac{\textstyle R_z\kb\phi\phi R_z}{\textstyle 1+\ip{\phi}{R_z\,\phi}}.
\end{eqnarray}
The first line results from writing $-\kb\phi\phi=(Q-z)-(Q'-z)$. This line is applied to the vector $\phi$ in the second line, and solved for $R'_z\phi$. This is then reinserted into the first line.

Now for (A) we need to show that $Q$ and $Q'$ have the same spectrum: absolutely continuous and equal to $\Rl$. For this we use the observation \cite[Thm.~XIII.20]{RSimon4} that for $Q'$ to have purely absolutely continuous spectrum it is sufficient that the matrix elements
$|\ip\psi{R'_{x+i\varepsilon}\psi}|$ be bounded as $\varepsilon\to0$, for some dense set of vectors and uniformly over intervals in $x$. Now from the resolvent formula \eqref{Krein} we see that this can be guaranteed by corresponding properties of the resolvent $R_z$ of $Q$.
Matrix elements of the resolvent can be rewritten as
\begin{equation}\label{matelres}
  \ip{\phi}{R_z\,\psi}=i\int_0^\infty dk\ e^{izk} \int_0^\infty dx\ e^{-ixk}\overline{\phi(x)}\psi(x)
\end{equation}
Now suppose that $\phi,\psi$ are polynomials times a Gaussian function. Then so is $\overline\phi\psi$ and its Fourier transform, which is the $x$-integral in the above expression. This makes the $k$-integral uniformly bounded for $z$ with small imaginary part. By multiplying $\phi$ with a suitable factor we can thus guarantee that
\begin{equation}\label{phibound}
  \left|\ip{\phi}{R_z\,\phi}\right|<1-\varepsilon
\end{equation}
with $\varepsilon>0$, uniformly for $z\in\Cx\setminus\Rl$ with small imaginary part. In fact, this is also necessary to exclude poles of the resolvent and hence eigenvalues. Furthermore, for a dense set of $\psi$, the matrix elements $ \ip{\phi}{R_z\,\psi}$ will also be bounded, i.e., we can conclude that $Q'$ has purely absolutely continuous spectrum. But then, because $\kb\phi\phi$ is, in particular, a trace class operator, the Kato-Rosenblum Theorem \cite[X.\S4~Thm4.4]{Kato} asserts that the absolutely continuous subspaces are unitarily equivalent. This proves claim (A).

Regarding claim (B), let us first establish the conditions that, for some initial state vector $\psi$, the probability distributions for $Q$ and $Q'$ coincide. Since by the resolvent equation products of resolvents can be converted to linear combinations, the span of the functions $E\mapsto1/(E-z)$  is a *-algebra which is dense in the set of all functions of $E$ vanishing at infinity. Therefore, our aim is equivalently formulated as finding criteria so that $\ip\psi{R_z\psi}=\ip\psi{R'_z\psi}$ for all $z$. By \eqref{Krein} this is equivalent to the product $ \ip{\psi}{R_z\,\phi} \ip{\phi}{R_z\,\psi}$ vanishing identically for $z\notin\Rl$. Since at least one of these analytic factors has to have accumulating zeros in the upper half plane, one factor has to vanish identically on the upper half plane. The other factor then automatically vanishes on the lower half plane. Now the vanishing of
\begin{equation}\label{Hardy0}
   \ip{\psi}{R_z\,\phi}=\int\frac{\overline{\psi(x)}\phi(x)}{x-z}\, dx
\end{equation}
for $\im(z)>0$ means that the ${\mathcal L}^1$-function $\overline{\psi}\phi$ is Hardy class, i.e., its Fourier transform vanishes on a half line. This happens sometimes (for example, when $\phi$ and $\overline\psi$ are both Hardy class), but (B) claims that for suitably chosen $\phi$ it never does.

Indeed, if the subspace generated by all $R_z\phi$ with $\im z>0$ is dense, we can find no vector $\psi$ such that \eqref{Hardy0} vanishes on a half plane. As a concrete example, let us again take a Gaussian $\phi$. We claim that, for any polynomial $p$, the function $p(x)\phi(x)$ is in the closed span of the vectors $R_z\phi$ with $\im z>0$. Since the set of functions $p\phi$ is dense in Hilbert space, this will prove claim (B). Given a polynomial $p$, choose some $z_\alpha$ with $\im z_\alpha>0$, namely at least as many as the degree of $p$, and form the function
\begin{equation}\label{partfrac}
  \tilde p(x)=p(x)\prod_\alpha \frac{z_\alpha}{x-z_\alpha}\ .
\end{equation}
By partial fraction decomposition this function can be written as a linear combination of the $1/(x-z_\alpha)$, so $\tilde p\phi$ is in the linear hull of the vectors $R_{z_\alpha}\phi$. On the other hand, taking $z_\alpha\to\infty$ in $\tilde p$ returns $p$, so $\tilde p\phi\to p\phi$ by dominated convergence. ---

\bibliography{UR}

\end{document}